\documentclass[fleqn,10pt]{wlscirep}
\usepackage[utf8]{inputenc}
\usepackage[T1]{fontenc}
\usepackage{soul}
\usepackage{comment}
\usepackage{hhline}
\usepackage{multirow}

\newcommand{\vo}{VO$_2$\:}

\title{Machine learning strategies to predict late adverse effects in childhood acute lymphoblastic leukemia survivors}

\author[1,*]{Nicolas Raymond}
\author[2,3]{Maxime Caru}
\author[1]{Hakima Laribi}
\author[1]{Mehdi Mitiche}
\author[4]{Valérie Marcil}
\author[5]{Maja Krajinovic}
\author[6]{Daniel Curnier}
\author[4]{Daniel Sinnett}
\author[1]{Martin Vallières}
\affil[1]{Department of Computer Science, Université de Sherbrooke, Sherbrooke, Canada}
\affil[2]{Division of Hematology and Oncology, Department of Pediatrics, Penn State College of Medicine, Hershey, PA, USA}
\affil[3]{Department of Public Health Sciences, Penn State College of Medicine, Hershey, PA, USA}
\affil[4]{Research Center, Sainte Justine University Health Center, Department of Nutrition, Université de Montréal, Montreal, Canada}
\affil[5]{Research Center, Sainte Justine University Health Center, Department of Pediatrics, Université de Montréal, Montreal, Canada}
\affil[6]{Research Center, Sainte Justine University Health Center, School of Kinesiology and Physical Activity Sciences, Faculty of Medicine, Université de Montréal, Montreal, Canada}

\affil[*]{nicolas.raymond2@usherbrooke.ca}

\begin{abstract}
Acute lymphoblastic leukemia is the most frequent pediatric cancer. Approximately two third of survivors develop one or more health complications known as late adverse effects following their treatments. The existing measures offered to patients during their follow-up visits to the hospital are rather standardized for all childhood cancer survivors and not necessarily personalized for childhood ALL survivors. As a result, late adverse effects may be underdiagnosed and, in most cases, only taken care of following their appearance. Thus, it is necessary to predict these treatment-related conditions earlier in order to prevent them and enhance the survivors’ health. Multiple studies have investigated the development of late adverse effects prediction tools to offer better personalized follow-up methods. However, no solution integrated the usage of neural networks to date. In this work, we developed graph-based parameters-efficient neural networks and promoted their interpretability with multiple post-hoc analyses. We first proposed a new disease-specific \vo peak prediction model that does not require patients to participate to a physical function test (e.g., 6-minute walk test) and further created an obesity prediction model using clinical variables that are available from the end of childhood ALL treatment as well as genomic variables. Our solutions were able to achieve better performance than linear and tree-based models on small cohorts of patients ($\leq$ 223) for both tasks.
\end{abstract}
\begin{document}

\flushbottom
\maketitle
%
%
\thispagestyle{empty}

\section*{Introduction}

Childhood acute lymphoblastic leukemia (ALL) is the most frequently diagnosed type of cancer in children \cite{Lemay:2019}. The 5-year relative survival rate is currently above 90\% \cite{Hunger:2012}. Nevertheless, approximately two thirds of childhood ALL survivors will present one or more health complications \cite{Nathan:2009} known as late adverse effects (LAEs). The LAEs are rather resulting from the treatment (e.g., exposure to chemotherapy, cranial radiation therapy) than the cancer itself \cite{Nathan:2009}. The existing follow-up measures, used in clinical settings and offered to patients during their visits to the hospital, are rather standardized for all childhood cancer survivors and not necessarily personalized for childhood ALL survivors \cite{Hudson:2021}. As a results, LAEs may be underdiagnosed, and in most cases, only taken care of once they have already appeared in adulthood. Thus, it is necessary to predict these treatments related conditions earlier in order to prevent them and enhance the survivors' health. 
 
Between 2013 and 2016, 246 childhood ALL survivors have participated to a series of clinical, physiological, biological and genetic evaluations as part of the PETALE study \cite{Marcoux:2017}. The main goal was to pinpoint predictive clinical, genetic and biochemical biomarkers that are relevant to establish personalized intervention plans to reduce LAEs prevalence, while providing knowledge for the improvement of follow-up methods \cite{Marcoux:2017}. 
 
Using the valuable data acquired from the PETALE cohort, efforts have been made towards the development of better personalized follow-up methods \cite{Labonte:2020, England:2017, Morel:2018, Nadeau:2019, Caubet:2019, Caru:2019}. As an example, an equation based on a linear regression was specifically developed to estimate the maximal oxygen consumption (i.e., \vo peak) in childhood ALL survivors following a 6-minute walk test (6MWT) \cite{Labonte:2020}. The \vo peak is an excellent predictor of cardiac health in patients with cancer and is recognized as the gold standard in exercise physiology to measure patients' cardiorespiratory fitness \cite{Smart:2013}, which plays an important role towards the prevention of LAEs in childhood ALL survivors \cite{Lemay:2019}. However, the direct measurement of the \vo peak, which is usually done by performing a maximal cardiopulmonary exercise test (CPET), is not an optimal solution in clinical settings due to financial and time constraints. Therefore, there is an interest in using a walking test (e.g., 6MWT) when access to comprehensive testing is limited (e.g., CPET) \cite{Mizrahi:2020}. Moreover, it has been shown that using a disease-specific \vo peak equation from the 6MWT provides a robust tool to estimate the patient's cardiorespiratory fitness with lower costs \cite{Mizrahi:2020}.

More recently, it has also been suggested that childhood ALL survivors' cardiorespiratory fitness is associated with specific trainability genes \cite{Caru:2019}, highlighting the potential impact of some genetic variants in the prediction of \vo peak. Another study investigated the association between genetic variants (i.e., single nucleotide polymorphisms) and cardiometabolic LAEs (e.g., obesity, dyslipidemia, hypertension) in childhood ALL survivors \cite{England:2017}. The single nucleotide polymorphisms (SNPs) were grouped according to their associated cardiometabolic conditions and further analyzed with eight other biological and treatment-related variables using a logistic regression. The authors found that multiple common and rare variants were independently associated with cardiometabolic conditions, such as dyslipidemia, insulin resistance and obesity \cite{England:2017}. They also suggested that these associations should be considered as indicators for the early assessment of these LAEs. This is an important aspect to take into consideration since, in the PETALE study, 41.8\% of childhood ALL survivors had dyslipidemia, 33\% were obese and 18.5\% had an insulin resistance \cite{England:2017}.

In medical contexts, simple models such as linear regression and logistic regression are often favored over more complex machine learning approaches (e.g., deep learning models) due to their ability of being easily interpreted \cite{Lundberg:2017, Fan:2021}. Moreover, due to their modest number of parameters to optimize (i.e., reduced capacity), simple models are less inclined to overfit on small training datasets and consequently have poor generalization performance. Hence, these models are well adapted to a clinical context with a small cohort of patients. However, more sophisticated model architectures (i.e., neural networks) have lately achieved better results in the prediction of clinical events using data from electronic health records \cite{Choi:2016, Ma:2017}. Interpretability of neural networks has also been the subject of many studies over the last years \cite{Fan:2021, Zhang:2021}. Post-hoc methods have been investigated to get insights about a neural network's behavior following its training. For example, recent work motivated the usage of attention mechanisms within their models to help depict the decision-making process behind individual samples \cite{Ma:2017, Arik:2021}. Model-agnostic techniques exist as well to compare and visualize features within a layer of a neural network \cite{Fan:2021, Maaten:2008, Mcinnes:2018}. On the other hand, interpretability of neural networks can also be strengthened a priori via the design of their architectures by including components with specific functionalities \cite{Fan:2021}. In particular, some studies explicitly integrated graph-based architectures (i.e., graph neural networks) to leverage the importance of the similarity between patients to solve a prediction task \cite{Lu:2021, Liu:2020}. 

\begin{figure}[ht!]
\centering
\includegraphics[scale=0.62]{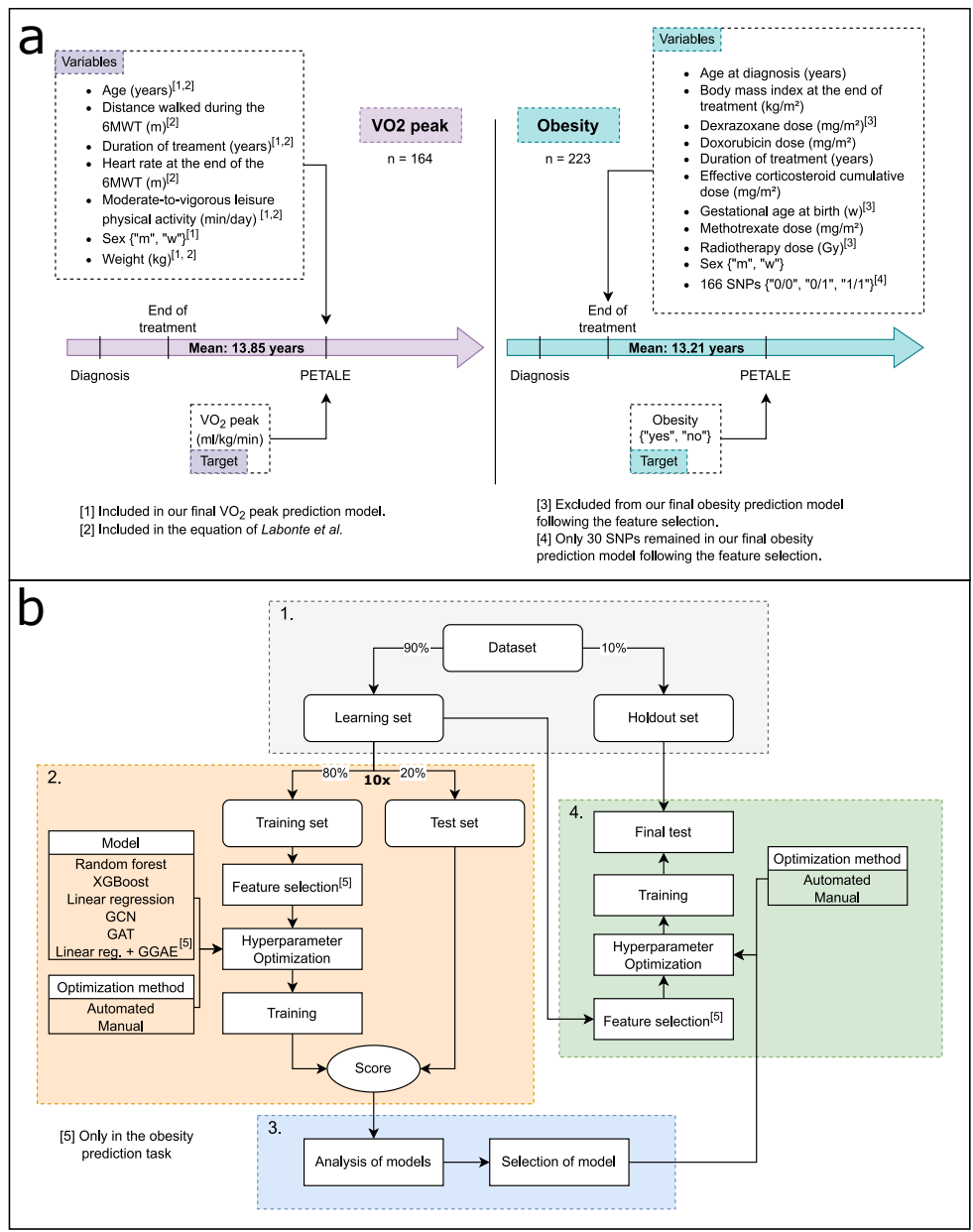}
\caption[Experimental setup]{Experimental setup. (\textbf{a}) Prediction tasks viewed using childhood ALL treatment timeline. On the left, \vo peak is predicted using variables measured on the same day. On the right, obesity is predicted using variables available at the end of childhood ALL treatment. (\textbf{b}) Experiment workflow. (1) Separation of the dataset into a \textit{learning set} and an \textit{holdout set}. (2) Evaluation of the models using random stratified subsampling with 10 splits. (3) Comparison of the models. (4) Final evaluation of the selected model on the \textit{holdout set}. Additional details are available in \textit{Experimental setup} section of Methods.}
\label{fig:experimental_setup}
\end{figure}

In this work, our main goal was to design neural networks for the prevention of LAES in childhood ALL survivors population. An overview of the prediction tasks and the experimental setup considered in this study is presented in figure~\ref{fig:experimental_setup}. We hypothesized that parameters-efficient neural networks could achieve better prediction performance than linear and tree-based models on small cohorts of patients but should require rigorous post-hoc analyses to provide interpretability of their behaviors. Especially, we believed that graph-based architectures would lead to the best results since they can benefit from the links between patients of the cohorts instead of treating each of them separately. We also suggested that the inclusion of genomic variables would be beneficial in the creation of early LAEs prediction model. Towards our goal, we first proposed a new disease-specific \vo peak prediction model that does not require patients to participate to a physical function test (e.g., 6MWT); even if it has some advantages over the cardiopulmonary exercise test, the 6MWT still requires time and human resources. We further created an obesity prediction model using clinical variables that are available from the end of childhood ALL treatment as well as genomic variables. Overall, our results suggest that neural networks can outperform simple models to predict LAEs in small cohorts of ALL survivors. The conducted post-hoc analyses including the visualization of features within specific layers and the visualization of attention maps also demonstrated their usefulness to provide a general understanding of the behaviors of the models.

\section*{Results}
 
\subsection*{Modeling the maximal oxygen consumption}

\subsubsection*{Construction of the prediction model}   
We developed a new regression model to estimate the \vo peak (mL/kg/min) in childhood ALL survivors. As opposed to the equation from Labonté \textit{et al.} \cite{Labonte:2020}, presented in Supplementary Background, we omitted the usage of variables related to the 6MWT but included the sex variable considering that it has an impact on the \vo peak \cite{Santisteban:2022}. Overall, we considered the age (years), the duration of treatment (DT) (years), the moderate-to-vigorous leisure physical activity (MVLPA) (min/day), the sex and the weight (kg) as our observed variables (Fig. \ref{fig:experimental_setup}\textbf{a}).

Recent works proposing the usage of Graph Neural Networks (GNNs) to improve prediction performance on datasets with no pre-established graph structure \cite{Chen:2020, Fatemi:2021, Qian:2021} motivated us to create an oriented graph from our dataset and build a model by combining the Jumping Knowledge Networks framework \cite{Xu:2018} to a Graph Attention Network (GAT) \cite{Velickovic:2018}. Instead of considering survivors individually, our model captures information from their neighborhood (i.e., the set of survivors connected to them by an oriented edge) when it calculates their predictions. Precisely, a targeted survivor encapsulates the information from his surrounding by calculating a weighted average of his neighbors' standardized features and by applying a transformation to the resulting vector. The weight attributed to each neighbor during the calculation of the weighted average is determined by the attention mechanism of the GAT. Both the attention mechanism and the transformation are parameterized functions for which the parameters are learnt during the training of the model. The vector resulting from the transformation is then concatenated to the initial standardized features to create an enriched representation of the survivor (i.e., an embedding).  It follows that the \vo peak of a survivor is estimated with a linear combination of the components within his embedding.

\subsubsection*{Construction of the graph}

We created the oriented graph structure connecting the childhood ALL survivors of our dataset using their attributes. To guide the attention mechanism towards a subset of connections with intuitively more potential to help with the regression task, we restricted the number of oriented edges pointing at each survivor (i.e., node) by selecting only their 10 nearest neighbors of the same sex. The similarity between survivors was determined using the Euclidean distance based on the numerical features. A self-connection was also added to each node so they could be part of their own neighborhood. The survivors from the \textit{holdout set} (Fig. \ref{fig:experimental_setup}\textbf{b}) were not allowed to be connected to each other in order for our experiment to be representative of a real clinical context where each new incoming survivor can only be connected to others that have already been observed (i.e., that are already in the graph).

\subsubsection*{Performance of the prediction model}
\begin{figure}[ht!]
\centering
\includegraphics[scale=0.60]{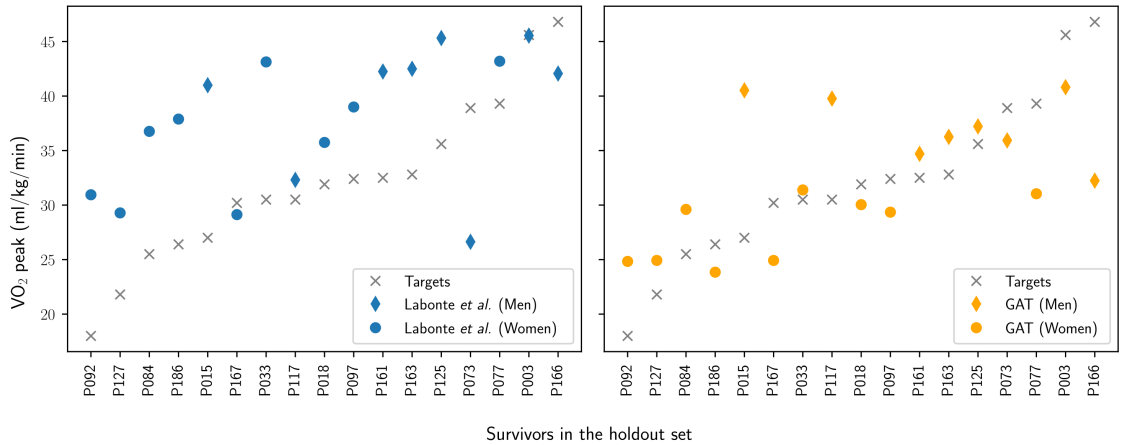}
\caption[Predictions of the equation from Labonté \textit{et al.} \cite{Labonte:2020} and the new \vo peak prediction model in the \textit{holdout set}.]{Predictions of the equation from Labonté \textit{et al.} \cite{Labonte:2020} and the new \vo peak prediction model in the \textit{holdout set}. On the left, the last established equation overestimates the \vo peak of the survivors. On the right, the new model based on a GAT architecture provides predictions that are closer to the targets.}
\label{fig:vo2_pred}
\end{figure}

We compared our new \vo peak prediction model to the equation from Labonté \textit{et al.} \cite{Labonte:2020} by measuring the root-mean-square error (RMSE), the mean absolute error (MAE), the Pearson correlation coefficient (PCC) and the concordance index (C-index) associated to the predictions of both models in the \textit{holdout set}. Except for the concordance index, our model shows an improvement against the equation from Labonté \textit{et al.} \cite{Labonte:2020} (\textit{final test} section of table \ref{tab:vo2}). In figure \ref{fig:vo2_pred}, we can see that the equation from Labonté \textit{et al.} \cite{Labonte:2020} is overestimating the \vo peak of childhood ALL survivors while our model is closer to the real observed values (i.e., targets). Moreover, our model does not rely on any measurement acquired from a 6MWT.

The results that led to the selection of our final model (i.e., GAT) are presented in the top section (i.e., \textit{evaluation of models}) of table \ref{tab:vo2}. The \textit{evaluation of models} consists of the second phase of our experimental setup (Fig. \ref{fig:experimental_setup}\textbf{b} box 2), in which a random forest, XGBoost \cite{Chen:2016}, a linear regression, a multi-layer Perceptron (MLP) and two GNNs (GCN \cite{Kipf:2017} and GAT \cite{Velickovic:2018}) combined with the Jumping Knowledge Networks framework \cite{Xu:2018} are compared against each other. Additional results associated to this experiment phase are available in Supplementary Table S13. Further details about the evaluation procedure and the description of the models are available in \textit{Experimental setup} and \textit{Models} subsections of Methods section respectively.

\begin{table}[ht!]
\centering
\resizebox{0.95\columnwidth}{!}{
\begin{tabular}{cc|cccc|c}
\cline{3-6}
                                                      &                            & \multicolumn{4}{c|}{Metric}                                                                             & \multicolumn{1}{l}{}                          \\ \hline
\multicolumn{1}{|c|}{Experiment phase}                & Model                      & \multicolumn{1}{c|}{RMSE} & \multicolumn{1}{c|}{MAE} & \multicolumn{1}{c|}{PCC} & C-index               & \multicolumn{1}{l|}{HPs optimization} \\ \hline \hline
\multicolumn{1}{|c|}{} & \textbf{Labonté \textit{et al.} \cite{Labonte:2020}}    & 7.95 $\pm$ 0.63              & 6.59 $\pm$ 0.56             & 0.77 $\pm$ 0.05             & 0.79 $\pm$ 0.03          & \multicolumn{1}{c|}{-}                        \\
\multicolumn{1}{|c|}{}                                & \textbf{Random forest}     & 5.44 $\pm$ 0.74              & 4.28 $\pm$ 0.54             & 0.79 $\pm$ 0.05             & \textbf{0.81 $\pm$ 0.02} & \multicolumn{1}{c|}{Manual}                \\
\multicolumn{1}{|c|}{}                                & \textbf{XGBoost}           & 5.39 $\pm$ 0.86              & \textbf{4.12 $\pm$ 0.68}    & 0.79 $\pm$ 0.06             & 0.81 $\pm$ 0.04          & \multicolumn{1}{c|}{Automated}                \\
\multicolumn{1}{|c|}{Evaluation of models}                                & \textbf{Linear regression} & 5.38 $\pm$ 0.70              & 4.22 $\pm$ 0.57             & 0.79 $\pm$ 0.07             & 0.80 $\pm$ 0.04          & \multicolumn{1}{c|}{Automated}                \\
\multicolumn{1}{|c|}{(\textit{learning set})}                                & \textbf{MLP}               & 5.76 $\pm$ 0.74              & 4.42 $\pm$ 0.54             & 0.77 $\pm$ 0.06             & 0.80 $\pm$ 0.03          & \multicolumn{1}{c|}{Automated}                   \\
\multicolumn{1}{|c|}{}                                & \textbf{GCN}               & 5.42 $\pm$ 0.86              & 4.14 $\pm$ 0.60             & 0.79 $\pm$ 0.07             & 0.81 $\pm$ 0.03          & \multicolumn{1}{c|}{Manual}                   \\
\multicolumn{1}{|c|}{}                                & \textbf{GAT}               & \textbf{5.34 $\pm$ 0.80}      & 4.13 $\pm$ 0.59             & \textbf{0.80 $\pm$ 0.07}    & 0.81 $\pm$ 0.03          & \multicolumn{1}{c|}{Manual}                   \\ \hline \hline
\multicolumn{1}{|c|}{Final test}     & \textbf{Labonté \textit{et al.} \cite{Labonte:2020}}    & \multicolumn{1}{c}{8.98}  & 7.84                     & 0.47                     & \textbf{0.72}                  & \multicolumn{1}{c|}{-}                        \\
\multicolumn{1}{|c|}{(\textit{holdout set})}                                & \textbf{GAT}               & \multicolumn{1}{c}{\textbf{6.51}  }           & \textbf{5.20}            & \textbf{0.52}            & 0.70       & \multicolumn{1}{c|}{Manual}                   \\ \hline
\end{tabular}
}
\caption[Performance of the models for the \vo peak prediction task]{Performance of the models for the \vo peak prediction task. Models are compared using the root-mean-square
error (RMSE), the mean absolute error (MAE), the Pearson correlation coefficient (PCC) and the concordance index (C-index). The results reported in the top section (i.e., \textit{evaluation of models}) refer to the \textit{mean $\pm$ standard deviation} obtained in the second part of our experimental setup (Fig. \ref{fig:experimental_setup}\textbf{b} box 2). The scores recorded following the predictions made by the selected model (i.e., GAT) and the equation from Labonté \textit{et al.} \cite{Labonte:2020} in the \textit{holdout set} are presented in the \textit{final test} section. The \textit{HPs optimization} column indicates if the scores were acquired with hyperparameter values that were manually selected or found by an automated hyperparameter optimization algorithm. \textbf{HP}: hyperparameter.} 
\label{tab:vo2}
\end{table}

\subsubsection*{Analysis of the prediction model}
\begin{figure}[ht!]
\centering
\includegraphics[scale=0.70]{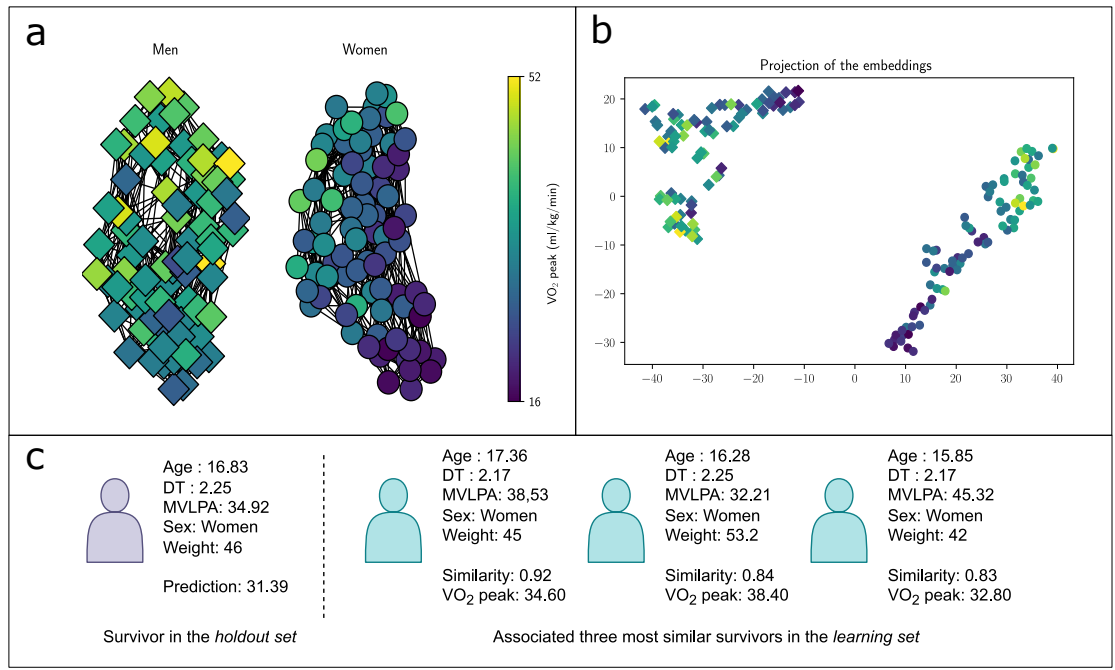}
\caption[Analysis of the \vo peak model]{Analysis of the \vo peak model. (\textbf{a}) Graph constructed for the \vo peak regression task. Connected men (diamonds) and women (circles) present similar \vo peak levels. The self-connections were omitted for visualisation purpose. (\textbf{b}) Projection of the embeddings in a 2D space using t-SNE \cite{Maaten:2008}. Men and women present two distinct groups where close survivors usually share similar \vo peaks. (\textbf{c}) Comparison between the profile of a survivor in the \textit{holdout set} and the associated three most similar survivors in the \textit{learning set} according to the embeddings learnt by the model.} 
\label{fig:vo2_analyses}
\end{figure}

We first visualized the oriented graph structure connecting the childhood ALL survivors of our dataset (Fig. \ref{fig:vo2_analyses}\textbf{a}). The resulting graph showed that survivors sharing connections had similar \vo peak values while supporting the fact that women in the childhood ALL survivors population have generally lower \vo peak values than men, as it is already observed in the non-survivors population \cite{Santisteban:2022}.

We further projected the embeddings learnt by our model in a 2D space using t-SNE \cite{Maaten:2008} in order to have a better understanding of the model's behavior (Fig. \ref{fig:vo2_analyses}\textbf{b}). The projection suggests that our model can learn embeddings that group survivors of the same sex while generally keeping them closer when they share similar \vo peak values. Considering that a shorter distance between two survivors' embeddings generally means that they have closer \vo peak values, we can help a clinician to validate the potential of a new prediction made by our model by comparing the profile of the survivor associated to the predicted value to the profiles of the associated most similar survivors for which we already know the \vo peak values. For example, in figure \ref{fig:vo2_analyses}\textbf{c}, we compared a survivor in the \textit{holdout set} with the associated three most similar survivors in the \textit{learning set}. In this case, the prediction made by the model seems legitimate since the closest patients in the \textit{learning set} have comparable attributes and \vo peak values. Note that the similarity measure in figure \ref{fig:vo2_analyses}\textbf{c} is based on a weighted Euclidean distance between the embeddings, that is, $similarity = 1/(1 + distance)$. The weight attributed to each dimension of the embeddings during the Euclidean distance calculation corresponds to the absolute value of the weight associated to the same dimension in the last linear layer of the model.

\subsection*{Modeling the obesity}

\subsubsection*{Construction of the prediction model}
We developed a new model for the early obesity prediction in childhood ALL survivors population. To make our model available to use directly at the end of the childhood ALL treatment, we considered the age at diagnosis (years), the body mass index (BMI) at the end of treatment (kg/m$^2$), the doxorubicin dose received (mg/m$^2$), the duration of treatment (DT) (years), the effective corticosteroid cumulative dose received (mg/m$^2$), the methotrexate dose received (mg/m$^2$), the sex and 30 SNPs as our observed variables (Fig. \ref{fig:experimental_setup}\textbf{a}). The SNPs are categorical variables that share the three following modalities: homozygous for the reference allele ("0/0"), heterozygous (i.e., one chromosome with the reference allele and the other with the alternate) ("0/1") or homozygous for the alternate allele ("1/1"). All variables mentioned above were kept following a feature selection process (see \textit{Feature selection} in Methods and Fig. \ref{fig:experimental_setup}\textbf{b}-4). The non-genomic variables excluded are shown in Supplementary Tables S4-S9.

The obesity status of a single individual can vary according to the measure used (e.g., body mass index (BMI), total body fat percentage (TBF), waist circumference) and the specific cut-off value associated to it, which can evolve according to guidelines. Thus, we trained our model to directly predict the future TBF of survivors. This way, any cut-off value can be further applied to evaluate if a survivor will be obese or not based on the predicted value. It allows our model to be independent from any cut-off value and consequently ensures that it stays operational with time. In our dataset, the time elapsed between the end of the treatment and the measurement of the TBF was on average 13.21 years (Fig. \ref{fig:experimental_setup}\textbf{a}).

\subsubsection*{Gene Graph Attention Encoder}
\begin{figure}[ht!]
\centering
\includegraphics[scale=0.65]{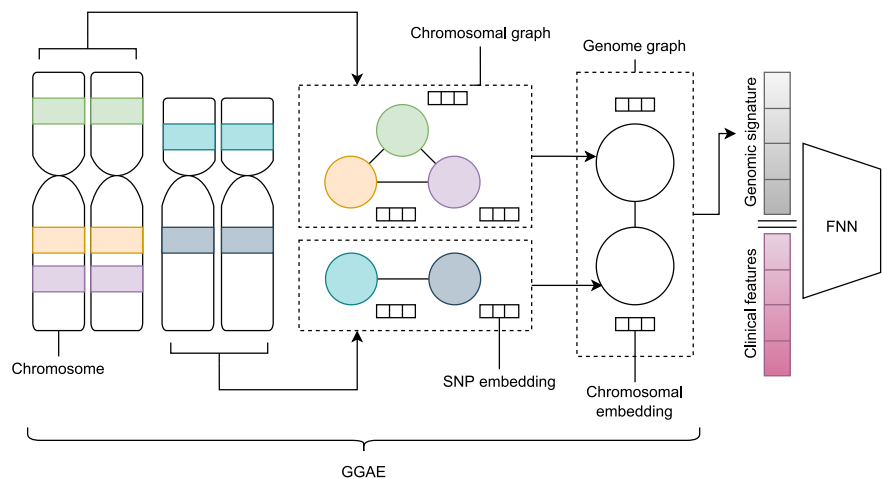}
\caption[The Gene Graph Attention Encoder (GGAE).]{The Gene Graph Attention Encoder (GGAE). In order to create the \textit{genomic signature}, the GGAE first interprets each chromosome pair as a complete graph (i.e., \textit{chromosomal graph}) where the nodes represent the observed SNPs associated to the pair. A real-valued vector is mapped to each node of each \textit{chromosomal graph} according to the SNP's category linked to the node (i.e., "0/0", "1/1" or "0/1"). We refer to each of these vectors as \textit{SNP embedding}. The GGAE further depict the whole genome as another complete graph (i.e., the \textit{genome graph}) where each node represents a pair of chromosomes. A real-valued vector (i.e., a \textit{chromosomal embedding}) is mapped again to each of these nodes by aggregating the \textit{SNP embeddings} of the associated \textit{chromosomal graph} (i.e., applying a readout function). Another readout function is finally applied to the \textit{genome graph} to create the \textit{genomic signature}.}
\label{fig:GGAE}
\end{figure}

We created a novel neural network architecture that efficiently uses the genomic data (i.e., the SNPs) for a regression task while considering the underlying structure of the genome. This new architecture called the Gene Graph Attention Encoder (GGAE) (Fig. \ref{fig:GGAE}) encodes the data from the SNPs in a low-dimensional vector named the \textit{genomic signature}. The latter is further concatenated to the other standardized clinical features to create a given patient embedding that can be used as an input to any feedforward neural network (FNN) architecture. The parameters of the GGAE and the subsequent connected FNN architecture are learned in an \textit{end-to-end} fashion during the training of the model. It follows that the model generates \textit{genomic signatures} that are specific  to the regression task. In our case, we used a simple linear regression (i.e., a linear layer) as the FNN part to reduce the number of parameters to optimize and allow the post-hoc analysis of the coefficients associated to the standardized clinical features. See section 2.2.6 of Supplementary Methods for further details on the architecture of the GGAE.

\subsubsection*{Performance of the prediction model}
\begin{figure}[ht!]
\centering
\includegraphics[scale=0.62]{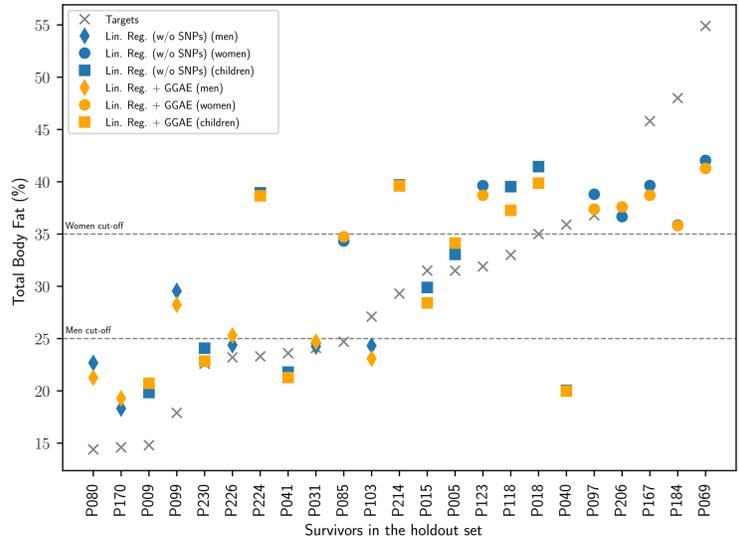}
\caption[Comparison of the linear regression with GGAE to the linear regression without the SNPs.]{Comparison of the linear regression with GGAE to the linear regression without the SNPs. The only obese man in the \textit{holdout set} was not identified by our model. Nonetheless, all obese women in the \textit{holdout set} were correctly identified.}
\label{fig:obesity}
\end{figure}

We compared the best model obtained with the inclusion of SNPs (linear regression + GGAE) to the best model obtained without the SNPs (linear regression) according to four different regression metrics (RMSE, MAE, PCC and C-index) and two binary classification metrics (sensitivity and specificity) (Table \ref{tab:obesity} \textit{final test} sections). The sensitivity and the specificity of the models were calculated following the categorization of the survivors as obese or not obese considering both the real TBFs and the model predictions with the cut-off values presented by Lemay \textit{et al.} \cite{Lemay:2019}: >25\% (men), >35\% (women) and >95th percentile (children). More precisely, for any child, the cut-off value was the 95th percentile measured from a sample of U.S children of the same sex and age group \cite{Ogden:2011}. The combination of the linear regression with the GGAE achieved the best scores on all metrics except for the specificity since it misclassified a non-obese survivor (P226) by 0.29 percentage point (Fig. \ref{fig:obesity}). Additional results associated to the \textit{evaluation of models} phase mentioned in the top sections of table \ref{tab:obesity} are available in Supplementary Tables~S14-S15.

\begin{table}[ht!]
\centering
\resizebox{\columnwidth}{!}{
\begin{tabular}{cc|llllll|c}
\cline{3-8}
                                                                                                           &                            & \multicolumn{6}{c|}{Metric}                                         \\ \hline
\multicolumn{1}{|c|}{Experiment phase}                                                                     & Model                      & \multicolumn{1}{c|}{RMSE}         & \multicolumn{1}{c|}{MAE}          & \multicolumn{1}{c|}{PCC}          & \multicolumn{1}{c|}{C-index}      & \multicolumn{1}{c|}{Sensitivity}  & \multicolumn{1}{c|}{Specificity}   & \multicolumn{1}{l|}{HPs optimization} \\ \hline \hline
\multicolumn{1}{|c|}{\multirow{6}{*}{\begin{tabular}[c]{@{}c@{}}Evaluation of models\\ (w/o SNPs) \\ (\textit{learning set})\end{tabular}}} & \textbf{Random forest}     & 8.97 $\pm$ 0.53                      & 7.36 $\pm$ 0.46                      & 0.65 $\pm$ 0.04            & 0.73 $\pm$ 0.02                      & \textbf{0.80 $\pm$ 0.06}             & 0.71 $\pm$ 0.12                       & \multicolumn{1}{c|}{Manual}          \\
\multicolumn{1}{|c|}{}                                                                                     & \textbf{XGBoost}           & 9.00 $\pm$ 0.50                      & \textbf{7.27 $\pm$ 0.41}             & 0.65 $\pm$ 0.04           & 0.74 $\pm$ 0.02                      & 0.77 $\pm$ 0.10                      & 0.71 $\pm$ 0.13                       & \multicolumn{1}{c|}{Automated}       \\
\multicolumn{1}{|c|}{}                                                                                     & \textbf{Linear regression} & \textbf{8.91 $\pm$ 0.52}             & 7.36 $\pm$ 0.45                      & \textbf{0.66 $\pm$ 0.05}                      & \textbf{0.75 $\pm$ 0.03}             & 0.77 $\pm$ 0.07                      & 0.70 $\pm$ 0.11                       & \multicolumn{1}{c|}{Automated}       \\
\multicolumn{1}{|c|}{}                                                                                     & \textbf{MLP}               & 9.07 $\pm$ 0.59                      & 7.27 $\pm$ 0.48                      & 0.64 $\pm$ 0.07                      & 0.74 $\pm$ 0.04                      & 0.68 $\pm$ 0.10                      & \textbf{0.76 $\pm$ 0.10}              & \multicolumn{1}{c|}{Manual}          \\
\multicolumn{1}{|c|}{}                                                                                     & \textbf{GCN}               & 9.03 $\pm$ 0.61     & 7.47 $\pm$ 0.53  & 0.65 $\pm$ 0.04    & 0.74 $\pm$ 0.02       & 0.76 $\pm$ 0.12            & 0.75 $\pm$ 0.10                       & \multicolumn{1}{c|}{Manual}          \\
\multicolumn{1}{|c|}{}                                                                                     & \textbf{GAT}               & 9.03 $\pm$ 0.52                      & 7.44 $\pm$ 0.52                      & 0.64 $\pm$ 0.04             & 0.73 $\pm$ 0.02                      & 0.72 $\pm$ 0.09                      & 0.75 $\pm$ 0.07                       & \multicolumn{1}{c|}{Manual}          \\ \hline \hline
\multicolumn{1}{|c|}{\multirow{7}{*}{\begin{tabular}[c]{@{}c@{}}Evaluation of models\\ (w/ SNPs) \\ (\textit{learning set})\end{tabular}}}  & \textbf{Random forest}     & 9.61 $\pm$ 0.71                      & 7.84 $\pm$ 0.67                      & 0.61 $\pm$ 0.04                      & 0.72 $\pm$ 0.02                      & \textbf{0.79 $\pm$ 0.09}             & 0.68 $\pm$ 0.10                       & \multicolumn{1}{c|}{Automated}       \\
\multicolumn{1}{|c|}{}                                                                                     & \textbf{XGBoost}           & 9.27 $\pm$ 0.58                      & 7.54 $\pm$ 0.47                      & 0.62 $\pm$ 0.04                      & 0.72 $\pm$ 0.02                      & 0.72 $\pm$ 0.12                      & 0.70 $\pm$ 0.10                       & \multicolumn{1}{c|}{Automated}       \\
\multicolumn{1}{|c|}{}                                                                                     & \textbf{Linear regression} & 9.56 $\pm$ 0.76                      & 7.95 $\pm$ 0.67                      & 0.58 $\pm$ 0.08                      & 0.70 $\pm$ 0.03                      & 0.71 $\pm$ 0.11                      & 0.69 $\pm$ 0.08                       & \multicolumn{1}{c|}{Automated}       \\
\multicolumn{1}{|c|}{}                                                                                     & \textbf{MLP}               & 9.73 $\pm$ 1.08                      & 7.80 $\pm$ 0.83                      & 0.60 $\pm$ 0.09                      & 0.72 $\pm$ 0.04                      & 0.68 $\pm$ 0.11                      & \textbf{0.74 $\pm$ 0.08}              & \multicolumn{1}{c|}{Automated}       \\
\multicolumn{1}{|c|}{}                                                                                     & \textbf{GCN}               & 9.82 $\pm$ 0.86                      & 7.92 $\pm$ 0.75                      & 0.58 $\pm$ 0.05                      & 0.70 $\pm$ 0.03                      & 0.65 $\pm$ 0.12                      & 0.68 $\pm$ 0.14                       & \multicolumn{1}{c|}{Automated}       \\
\multicolumn{1}{|c|}{}                                                                                     & \textbf{GAT}   &      9.64 $\pm$ 0.67              & 7.80 $\pm$ 0.63             & 0.59 $\pm$ 0.05             & 0.71 $\pm$ 0.02                 & 0.72 $\pm$ 0.14                     & 0.70 $\pm$ 0.11                                  & \multicolumn{1}{c|}{Automated}       \\
\multicolumn{1}{|c|}{}                                                                                     & \textbf{Lin. reg. + GGAE}  & \textbf{9.02 $\pm$ 0.57}             & \textbf{7.47 $\pm$ 0.55}             & \textbf{0.65 $\pm$ 0.05}             & \textbf{0.74 $\pm$ 0.02}             & 0.77 $\pm$ 0.08                      & 0.72 $\pm$ 0.12                       & \multicolumn{1}{c|}{Manual}          \\ \hline \hline
\multicolumn{1}{|c|}{\begin{tabular}[c]{@{}c@{}}Final test \\ (w/o SNPs) \\ (\textit{holdout set})\end{tabular}}                     & \textbf{Linear regression} & \multicolumn{1}{c}{7.96}          & \multicolumn{1}{c}{6.31}          & \multicolumn{1}{c}{\textbf{0.67}} & \multicolumn{1}{c}{0.75}          & \multicolumn{1}{c}{\textbf{0.83}} & \multicolumn{1}{c|}{\textbf{0.88}} & \multicolumn{1}{c|}{Manual}          \\ \cline{1-1}
\multicolumn{1}{|c|}{\begin{tabular}[c]{@{}c@{}}Final test\\ (w/ SNPs) \\ (\textit{holdout set})\end{tabular}}                       & \textbf{Lin. reg. + GGAE}  & \multicolumn{1}{c}{\textbf{7.87}} & \multicolumn{1}{c}{\textbf{6.27}} & \multicolumn{1}{c}{\textbf{0.67}} & \multicolumn{1}{c}{\textbf{0.76}} & \multicolumn{1}{c}{\textbf{0.83}} & \multicolumn{1}{c|}{0.82}          & \multicolumn{1}{c|}{Manual}          \\ \hline
\end{tabular}
}
\caption[Performance of the models for the obesity prediction task.]{Performance of the models for the obesity prediction task. Regression metrics are calculated using the predictions of TBF. Classification metrics are calculated considering the obesity class (obese or not obese) associated to each prediction following the application of the cut-off values. The results reported in the \textit{evaluation of models} sections refer to the \textit{mean} ± \textit{standard deviation} obtained in the second part of our experimental setup (Fig. \ref{fig:experimental_setup}\textbf{b} box 2). The scores achieved by the selected model without SNPs (linear regression) and the selected model with SNPs (linear regression + GGAE), in the \textit{holdout set}, are displayed in the \textit{final test} sections. The \textit{HPs optimization} column indicates if the scores were acquired with hyperparameter values that were manually selected or found by an automated hyperparameter optimization algorithm. \textbf{HP}: hyperparameter.}
\label{tab:obesity}
\end{table}

\subsection*{Analysis of the prediction model}
\subsubsection*{Attention mechanism}
\begin{figure}[ht!]
\centering
\includegraphics[scale=0.80]{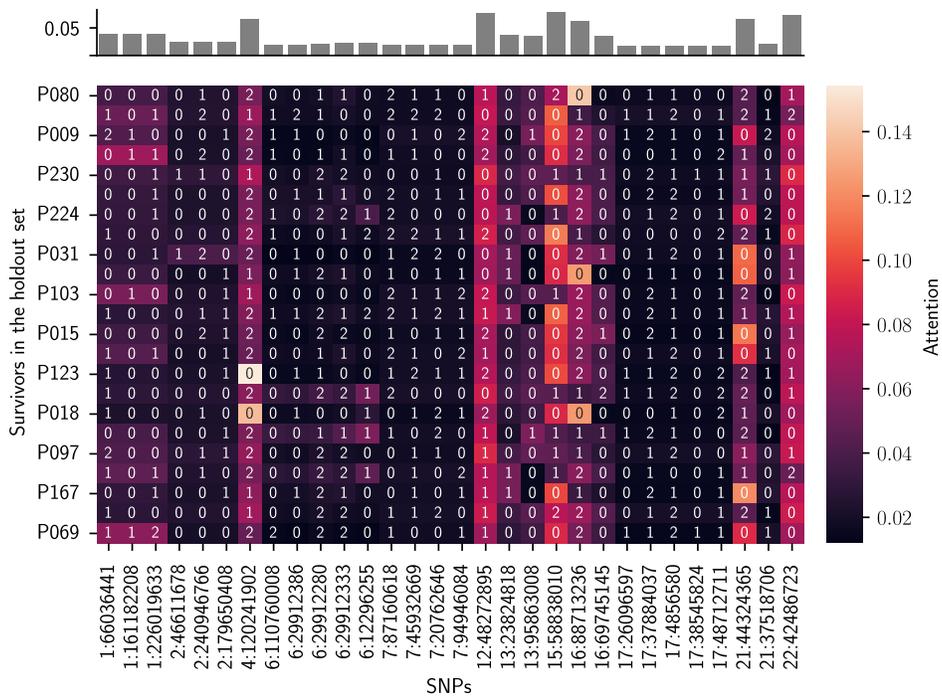}
\caption[SNPs attention heatmap]{SNPs attention heatmap. We display the attention score (in the range $[0, 1]$) given to each SNP by each survivor in the \textit{holdout set}. The survivors are sorted from the lowest TBF to the highest. The attention scores in each row are summing to 1.  Each part of the grid is annotated with a "0", a "1" or a "2" to show if the SNP of a survivor is respectively homozygous for the reference allele ("0/0"), heterozygous ("0/1") or homozygous for the alternate allele ("1/1"). A bar plot with the average attention score of each SNP is presented over the heatmap.}
\label{fig:heatmap}
\end{figure}

The attention mechanism in the GGAE demonstrated its capability to focus on different SNPs to make a prediction. The attention of the GGAE was mainly directed toward the SNPs 4:120241902, 12:48272895, 15:58838010, 16:88713262, 21:4432365 and 22:42486723 (Fig. \ref{fig:heatmap}). The SNPs 4:120241902, 15:58838010, 16:88713262 and 21:4432365 had higher attention scores when they were homozygous for the reference allele. Finally, the attention scores given to SNPs 12:48272895 and 22:42486723 were high for
any category.

\subsubsection*{Impact of the clinical features}
We analyzed the coefficients associated to the standardized clinical features and the intercepts related to each sex. The age at diagnosis (0.05), the BMI at the end of treatment (2.33), the doxorubicin dose received (0.65) and the effective corticosteroid cumulative dose received (0.44) were found to increase the prediction of the TBF. The obesity at the end of treatment was already identified as a prevalence factor of the obesity at the interview for the survivors in the PETALE cohort \cite{Levy:2017}. Additionally, corticosteroids have also been reported to increase the obesity risk in other studies \cite{Van:1995, Chow:2007}. Therefore, it is legitimate that positive coefficients are associated to the BMI at the end of treatment and the cumulative corticosteroids dose received. On the other hand, the duration of treatment (-0.32) and the methotrexate dose (-1.56) were found to decrease the prediction of the TBF. The intercepts calculated for both sex (men: 16.31, women: 30.35) support the fact that women have generally higher TBF than men, which is common in the non-childhood ALL survivors \cite{Karastergiou:2012}.

\section*{Discussion}
Over the years, efforts have been pursued towards the development of better personalized follow-up methods for childhood ALL survivors using data from the PETALE study \cite{Labonte:2020, England:2017, Morel:2018, Nadeau:2019, Caubet:2019, Caru:2019}. Other recent works have presented interesting results associated to the usage of neural networks in prediction tasks related to clinical contexts \cite{Choi:2016, Ma:2017}. However, until now, these machine learning approaches remained underexplored for the prediction of LAEs in childhood ALL survivors. In our work, we developed graph-based parameters-efficient neural networks for the LAEs prediction in childhood ALL survivors. In addition to contributing to precision medicine, our solutions constitute a promising avenue for the usage of artificial intelligence in clinical settings with restricted numbers of patients.

We first created a new disease-specific \vo peak prediction model based on a Graph Attention Network \cite{Velickovic:2018}. The \vo peak is the gold standard to measure the cardiorespiratory fitness \cite{Smart:2013}, which in turn is a key element for the prevention of LAEs such as obesity, cholesterol and depression \cite{Lemay:2019}. To use this type of neural network architecture and handle the \vo peak prediction task as a node regression problem, we created a graph structure with our dataset (Fig. \ref{fig:vo2_analyses}\textbf{a}). In addition to achieving better performance than the equation from Labonté \textit{et al.} \cite{Labonte:2020}, our model does not rely on a walking test (e.g., 6MWT). The removal of this constraint represents a strong advantage in the context of healthcare considering that the 6MWT requires time and financial resources. Moreover, with our new model, the \vo peak prediction becomes more accessible since all variables needed by the model can be self-reported by the survivors. Therefore, our model could be associated to an online survey that survivors would be requested to fill at different time points. The resulting predictions could be further analysed by an exercise physiologist with the support of an interface providing comparisons between the current patient and the most similar survivors for which the \vo peak is already known (Fig. \ref{fig:vo2_analyses}\textbf{c}). Even tough it obtained satisfying results, the \vo peak prediction model presented development challenges that should be further addressed in following works. The manual construction of an optimal graph structure represents a tedious task. Until now, we only explored solutions based on the calculation of distances between the observations of our dataset. However, even if these solutions are conceptually simple, they involve the selection of several additional hyperparameters such as the choice of a distance metric and the number of neighbors associated to each node. Machine learning methods enabling to simultaneously learn the graph structure relative to the data as well as the parameters of the model should be considered. Among the recent developments relevant to the subject, the \textit{Graph Convolutional Transformer} \cite{Choi:2020} is a an example of model that simulates the presence of an edge between any pair of observations within a dataset and learn how to calculate a weight for each of them. In other words, this model features a flexible mechanism allowing each node to determine the best candidates to be part of its neighborhood. Whilst such model's complexity is growing according to the number of nodes in a graph, it is a plausible solution to consider in future work given the small cohort size enrolled in our study. In addition to simplifying the construction of the graph, this approach could lead to the improvement of our current state-of-the-art solution.

We also proposed an obesity prediction model using clinical variables that are available at the end of childhood ALL treatment as well as genomic variables. In addition to showing promising result in the prediction of future obesity, our work presented a novel neural network architecture (i.e., the GGAE) that efficiently encodes the information associated to the SNPs (Fig. \ref{fig:GGAE}). Its design improve the modeling of genomic data while allowing to manage the obesity prediction task similarly to a graph classification problem. The attention map produced by the GGAE (Fig. \ref{fig:heatmap}) demonstrates the degree of follow-up personnalization pursued by our work. Not only it allows to generate hypothesis about the importance of certain SNPs regarding the survivors population in general, but it also enables to see the contribution of the allelic constituents within each individual. For example, the map produced for the \textit{holdout set} suggests that certain SNPs (i.e., 4:120241902, 12:48272895, 15:58838010, 16:88713262, 21:4432365 and 22:42486723) could be generally more relevant than the others for the prediction of the future TBF (Top of Fig. \ref{fig:heatmap}). Meanwhile, it also indicates on which SNPs the model was focusing the most according to each patient. Among others, we can mention the higher attention level given to the SNP 4:120241902 by patient P018 and P123 (Fig. \ref{fig:heatmap}). We acknowledge that the performance gain provided by the usage of SNPs through the GGAE was small (Fig. \ref{fig:obesity}) and therefore,  a study comparing the benefits and the cost of executing a whole exome sequencing should be conducted. Nonetheless, we consider the GGAE as an innovative solution for the integration of heterogeneous oncological data in parameters-efficient neural networks and we plan to further explore its potential in future work. It should be noted that the association between the SNP 12:48272895 (i.e., the VDR FokI polymorphism) and different obesity traits has already been investigated in multiple studies but results were found to be inconsistent \cite{Alathari:2020}.

We further highlighted limitations related to our study. First, only a small number of samples were available in the PETALE dataset. Therefore, the scores obtained in the \textit{holdout sets} related to both prediction tasks might not be fully representative of the future performance of our models on bigger and unseen datasets. Moreover, we hypothesize that the limited number of samples reduced the effectiveness of the automated hyperparameter optimization. More precisely, even though each set of hyperparameters selected by the algorithm was evaluated on multiple sub-samples, it is possible that their sizes were too small to provide valuable estimations of the hyperparameters' reliability. Second, all survivors of our datasets were from a monocentric cohort where individuals had only European origins. Hence, our findings may not translate to other ethnicity groups of childhood ALL survivors. Third, the concept of future is not clearly defined within the current design of the obesity prediction model. The time from the end of treatment could eventually be integrated as an additional variable to predict this LAE within a more precise time frame while potentially contributing to an increase of its accuracy \cite{Aaron:2019}. 

Next, we believe that future work could be separated in three different phases: (i) the validation of the current work using an external dataset; (ii) the application of the current work to other LAEs; (iii) the development of new architectures based on multi-task learning\cite{Ruder:2017}. During the validation phase, performance of the new models could be first tested on other cohorts of childhood ALL survivors with European origins. Tests could also be performed on cohorts with different ethnicity to acquire more information about the clinical settings in which our models are reliable. Additionally, investigation could be pursued concerning the possible association between the TBF and the SNPs that received higher attention scores by the GGAE. Except for the VDR FokI polymorphism, we did not find any work that reported relevant results regarding the link between these SNPs and the TBF. At last, investigations could also be conducted to further quantify the theoretical benefits of using GNNs with datasets that do not naturally have underlying graph structures. In the second phase, considering the promising results achieved in our work for the prediction of the \vo peak and the future TBF, the development of prediction models for other LAEs such as dyslipidemia and insulin resistance would be an interesting avenue to explore. In terms of the third phase, we hypothesize that, on small cohorts, neural networks designed for the simultaneous prediction of multiple LAEs via multi-task learning could provide better performance than neural networks built to produce a single output. This hypothesis follows the intuition that a neural network trained to predict LAEs within a same family (e.g., metabolic disorders) should benefit from the common underlying patterns linking the variables to each specific LAE, while being less vulnerable to overfitting since it has to learn parameters that help conjointly for the different tasks.

In conclusion, we demonstrated in our work that graph-based parameters-efficient neural networks can achieve better results than linear and tree-based models for prediction tasks in clinical contexts with small cohorts ($\leq$ 223 childhood ALL survivors). We also showed that an improvement of regression performance can be leveraged from the creation of graph-based architectures by either connecting patients of a dataset together or providing a better modeling of their individual information (e.g., genomic data). Additionally, we displayed that it is feasible to have a better understanding of the behaviors of these more complex machine learning solutions with post-hoc analysis methods such as the visualization of patients' embeddings and the study of attention maps. Overall, we strongly believe that the design of efficient model architectures and the achievement of post-hoc analyses are the key to increase the progress and the trust associated to the usage of machine learning with small cohorts in healthcare.

\section*{Methods}

\subsection*{Datasets}
All data was taken from the PETALE study. All participants of this study were survivors with European origins who have been diagnosed for childhood ALL between 1987 and 2010 before the age of 19 and were at least 5 years post-diagnosis (see the article from Marcoux \textit{et al.} \cite{Marcoux:2017} for a complete list of the eligibility criterion). Descriptive analyses of the datasets and procedures of their construction are presented in Supplementary Tables S1-S12 and Supplementary Figures S1-S2 respectively. The \vo peak dataset consisted of 164 survivors who reached a valid maximal oxygen consumption while performing a cardiopulmonary exercise test \cite{Labonte:2020}. 90\% of the survivors with a \vo peak under or equal to the median were women while 76\% of the survivors with a \vo peak over the median were men. The obesity dataset consisted of 223 survivors for which the TBF was measured by dual energy x-ray absorptiometry \cite{Lemay:2019}. 76\% of the survivors with a TBF under or equal to the median were men while 77\% of the survivors with a TBF over the median were women.

\subsection*{Experimental setup}
We developed a framework (Fig. \ref{fig:experimental_setup}\textbf{b}) to compare the performance of different models (see Models section) for the \vo peak and the obesity regression tasks. In this framework, 10\% of the dataset is first extracted using random stratified sampling (see Random stratified sampling section) to create a \textit{holdout set}. The \textit{holdout set} remains hidden until the final best model is selected and ready to be evaluated. The search of the best model is done using the 90\% of data left, which is referred to as the \textit{learning set}. The latter is divided 10 times into different training sets and test sets. Each test set is constituted of 20\% of the \textit{learning set} and is also extracted using random stratified sampling. For the early obesity problem, a selection of features is conducted on each of these data splits considering the data from the training set (see Feature selection section) to exclude variables that are not helpful for the prediction.

Each model is evaluated on these 10 data splits at least twice. The models are first evaluated considering manually selected sets of hyperparameter values. Then, models are evaluated using a set of hyperparameter values obtained from an automated hyperparameter optimization algorithm (see Hyperparameter optimization section). For each model evaluation, we save the empirical means and standard deviations of the metrics calculated on the test sets for further analyses. 
The model that achieved the best performance for the greatest number of metrics during one of its evaluations is kept as our final model. The best manually selected set of hyperparameter values, as well as the hyperparameters' search spaces used for the automated hyperparameter optimization of each model, are reported for each model in section 2.3.2 of Supplementary Methods.

The selected model is finally trained and evaluated twice on the \textit{learning set} and the \textit{holdout set} respectively. The first training is done using the best manually selected set of hyperparameter values and the second training is done using automated hyperparameter optimization. For the  early obesity problem, a selection of features is conducted beforehand on the \textit{learning set} to exclude variables that are not helpful for the prediction. The selected features are reported in the Results section.

Noteworthy, the model comparison step (Fig. \ref{fig:experimental_setup}\textbf{b} box 2) was conducted twice for the early obesity prediction task. We have first selected the best model by running the comparisons without the SNPs and then selected the best model considering the SNPs. Both models were finally evaluated on the \textit{holdout set} (Table \ref{tab:obesity}).

\subsection*{Models}
For each regression task and each set of variables tested, we evaluated the performance of a random forest, XGBoost \cite{Chen:2016}, a linear regression (trained with gradient descent), a multi-layer Perceptron (MLP) and two GNNs (GCN \cite{Kipf:2017} and GAT \cite{Velickovic:2018}) combined with the Jumping Knowledge Networks framework \cite{Xu:2018}. The linear regression with the GGAE was only evaluated for the obesity task with the set of variables including the SNPs. The random forest and the XGBoost implementations were taken respectively from \textit{scikit-learn} \cite{Pedregosa:2011} and \textit{xgboost} libraries. The other models were implemented using \textit{PyTorch} \cite{Paszke:2019} and \textit{DGL} \cite{Wang:2019} librairies. Each model had predetermined sets of values and search spaces associated to their hyperparameters (Supplementary Methods section 2.3.2). These sets were respectively used to execute the experiments without hyperparameter optimization and with hyperparameter optimization. The architectures of the models and the descriptions of their hyperparameters are available in sections 2.2 and 2.3.1 of Supplementary Methods respectively.

\subsection*{Hyperparameter optimization}
Hyperparameter optimization (Supplementary Fig. S3) was conducted by evaluating 200 sets of hyperparameter values sampled using the Tree-structured Parzen Estimator algorithm (TPE) \cite{Bergstra:2011}. Each set was evaluated on 10 \textit{internal training sets} and \textit{internal test sets} created by sub-dividing the training set (as well as the \textit{learning set}) using stratified random sampling (see Random stratified sampling section). The average RMSE observed in the 10 different \textit{internal test sets} was used to estimate the performance related to a set of hyperparameter values. The set of hyperparameter values associated to the lowest average RMSE was selected.  All the hyperparameter optimization process was executed using the \textit{optuna} \cite{Akiba:2019} library. The settings of the TPE algorithm are reported in Supplementary Tables S27-S28.

\subsection*{Random stratified sampling}
All test sets created (as well as the \textit{holdout sets} and the \textit{internal test sets}) were sampled using random stratified sampling. The stratification was performed each time on a temporary column combining sex and discretized versions of the targets (i.e., the \vo peak values or the TBFs depending on the regression task) based on the median in the complete dataset (i.e., the dataset before the extraction of the \textit{holdout set}). The temporary column had four modalities: women ($\leq$median), women (>median), men ($\leq$median) and men (>median).  The sex was considered in the stratification since we knew beforehand that it had an impact on the the \vo peak \cite{Santisteban:2022} and the TBF \cite{Karastergiou:2012} in the non-childhood ALL survivors population. 

Moreover, two additional criteria were established to verify if a test set (as well as a \textit{holdout set} or an \textit{internal test set}) was valid. The first criterion was that, for any test set sampled, the remaining dataset must contain any possible modality associated to the categorical variables. This criterion ensures that any categorical modality has been considered during the training of a model and can further be recognized during the evaluation of the same model on the test set. The second criterion was that the numerical values observed within any numerical column of the test set must not be further than 6 interquartile ranges away from the first and third quartiles of the same column in the remaining dataset. This criterion ensures that the numerical values in the test set lies in a similar region than the one see in the set used for the training.

\subsection*{Feature selection}
For each training set (as well as each \textit{learning set}), we trained 10 different random forests using the default hyperparameters of version 0.24.1 of \textit{scikit-learn}. We further extracted the feature importance calculated by each random forest for each feature. All the features with an average feature importance greater or equal to 0.01 were kept for the training. The selection of the clinical features and the genomic features (i.e., the SNPs) was done independently.

\subsection*{Data imputation and transformation}
For each pair of training and test sets created (as well as \textit{learning}/\textit{holdout} and \textit{internal training}/\textit{internal test} sets pairs), we imputed missing data in the numerical columns using the empirical means calculated with the observed data in the training set and imputed the missing data in the categorical columns using the modes of the observed data in the training set. Once imputed, transformation steps were applied to each pair of training and test sets. Numerical columns were reduced and centered using the empirical means and standard deviations calculated with the observed data in the training set. The modalities of each categorical column were changed for nominal encodings.

\subsection*{Graph construction}
The directed graphs considered to train and evaluate the GNNs were built by considering the attributes of the survivors. Especially, the oriented edges pointing at each survivor (i.e., node), were coming from the nodes of the $k$-nearest neighbors of the same sex. During the \textit{evaluation of models} phase (Fig. \ref{fig:experimental_setup}\textbf{b} box 2), values of $k$ of 4, 6, 8 and 10 were considered for the evaluation of each GNN model with manually selected hyperparameters (Supplementary Tables S16-S18). The value of $k$ associated to the best performance of a GNN, with manually selected hyperparameters, was further used during the automated hyperparameter optimization of the same model (Supplementary Tables S24-S25). The similarity between each survivor was calculated on all the standardized numerical features and categorical features excluding the sex. More precisely, we used the value 1/(1 + Euclidean distance) in cases where no categorical attributes were available and considered the cosine similarity otherwise. The categorical attributes were converted to one-hot encodings to calculate the cosine similarities. The similarity values were set as the weights of the edges for the GCN (Supplementary Methods section 2.2.4).

Survivors in the test sets (as well as the \textit{holdout sets} and the \textit{internal test sets}) were always excluded from each graph during the training of GNNs. Moreover, once added for the evaluation, survivors in the test sets were only allowed to have oriented edges coming from the nodes of the associated training graph. As an example, for each prediction task, during the execution of the second phase of the experimental setup (Fig. \ref{fig:experimental_setup}\textbf{b} box 2), training graphs were only constituted of patients in the training set while patients of the associated test sets were only added in the graphs for the evaluations after the training.

\subsection*{Ethics declarations}
All the analyses conducted for the PETALE study were compliant with the Declaration of Helsinki and approved by the Institutional Review Board of Sainte-Justine University Health Center. Written informed consent was obtained from study participants or parents/guardians.

\section*{Code Availability}
All software code allowing to run the experiments used
to produce all the results presented in this work is freely shared under the GNU General Public License v3.0 on the GitHub website at: \href{https://github.com/Rayn2402/ThePetaleProject}{https://github.com/Rayn2402/ThePetaleProject}.

\section*{Data Availability}
The datasets analysed during the current study are not publicly available for confidentiality purposes. However, randomly generated datasets with the same format as used in our experiments are publicly shared in our GitHub \href{https://github.com/Rayn2402/ThePetaleProject}{repository} to test the code implemented for this work.

\bibliography{article}

\section*{Acknowledgements}
This work was supported by the Canadian Institutes of Health Research (CIHR), in collaboration with C17 Council, Canadian Cancer Society (CCS), Cancer Research Society (CRS), Garron Family Cancer Centre at the Hospital for Sick Children, Ontario Institute for Cancer Research (OICR) and Pediatric Oncology Group of Ontario (POGO). VM is supported by the Fonds de recherche Québec-Santé. DS holds the research chair FKV in pediatric oncogenomics. MV acknowledges funding from the CIFAR AI Chairs program. 

\section*{Author contributions statement}
NR and MV conceived the study. NR wrote the manuscript and performed data analysis. NR and MM implemented the machine learning framework used to perform data analysis. HL generated random datasets to validate experiment reproducibility. MV supervised the study. MC, VM, MK, DC and DS contributed to the experimental design. All authors revised the manuscript.

\section*{Additional information}
\subsection*{Competing interests statement}
The author(s) declare no competing interests.

\end{document}


\begin{titlepage}
\begin{center}
\scshape\LARGE \textbf{SUPPLEMENTARY INFORMATION}
\end{center}

\vspace{3in}

\begin{center}
\scshape\LARGE \emph{Machine learning strategies to predict late adverse effects in childhood acute lymphoblastic leukemia survivors}
\end{center}

\vspace{3in}

\begin{center}
    AUTHORS:\\
Nicolas Raymond, Maxime Caru, Hakima Laribi, Mehdi Mitiche, Valérie Marcil, Maja Karjinovic, Daniel Curnier, Daniel Sinnett \& Martin Valli\`eres
\end{center}

\thispagestyle{empty}
\end{titlepage}
\setcounter{page}{2}

\tableofcontents
\newpage
\listoftables
\listoffigures
\newpage


\section{SUPPLEMENTARY RESULTS}
\subsection{Descriptive analyses of the datasets}
In this subsection we present tables with descriptive analyses of the complete dataset, the \textit{learning set} and the \textit{holdout set} related to each prediction task. The statistics associated to each numerical feature are presented into the format \textit{mean} (\textit{\acrshort{std}}) [\textit{min}, \textit{max}]. The statistics associated to each categorical feature (e.g., \acrshort{snp}) are presented into the format \textit{count} (\textit{percentage of group}). Note that the statistics are calculated without considering any missing value.

\begin{table}[ht!]
\centering
\resizebox{\columnwidth}{!}{
\begin{tabular}{c|ccc|}
\cline{2-4}
\multicolumn{1}{l|}{}                     & \multicolumn{3}{c|}{Group}                                                                                                                          \\ \hline
\multicolumn{1}{|c|}{Feature}            & \multicolumn{1}{c|}{All (n = 164)}                     & \multicolumn{1}{c|}{Women (n = 84)}                    & Men (n = 80)                      \\ \hline \hline
\multicolumn{1}{|c|}{Age (years)}         & \multicolumn{1}{c|}{22.40 (6.13) {[}11.33, 38.76{]}}   & \multicolumn{1}{c|}{22.50 (6.22) {[}11.33, 38.76{]}}   & 22.29 (6.07) {[}11.40, 36.92{]}   \\ \hline
\multicolumn{1}{|c|}{\acrshort{dt} (years)}          & \multicolumn{1}{c|}{2.13 (0.17) {[}1.17, 2.83{]}}      & \multicolumn{1}{c|}{2.16 (0.18) {[}1.17, 2.83{]}}      & 2.10 (0.16) {[}1.33, 2.41{]}      \\ \hline
\multicolumn{1}{|c|}{\acrshort{hrend} (bpm)} & \multicolumn{1}{c|}{149.37 (22.33) {[}75.00, 191.00{]}}  & \multicolumn{1}{c|}{144.87 (23.21) {[}75.00, 191.00{]}}  & \multicolumn{1}{c|}{153.92 (20.57) {[}96.00, 190.00{]}}  \\ \hline
\multicolumn{1}{|c|}{\acrshort{mvlpa} (min/day)}     & \multicolumn{1}{c|}{25.26 (25.03) {[}0.00, 149.18{]}}  & \multicolumn{1}{c|}{24.09 (25.66) {[}0.00, 149.18{]}}  & 26.49 (24.45) {[}0.00, 110.16{]}  \\ \hline
\multicolumn{1}{|c|}{Weight (kg)}         & \multicolumn{1}{c|}{66.75 (15.94) {[}40.50, 113.00{]}} & \multicolumn{1}{c|}{64.37 (16.66) {[}40.50, 113.00{]}} & 69.26 (14.83) {[}43.00, 108.00{]} \\ \hline
\multicolumn{1}{|c|}{\acrshort{6mwd} (m)}   & \multicolumn{1}{c|}{612.25 (78.47) {[}412.00, 873.00{]}} & \multicolumn{1}{c|}{637.53 (80.55) {[}412.00, 873.00{]}} & \multicolumn{1}{c|}{587.60 (68.32) {[}445.00, 755.00{]}} \\ \hline \hline
\multicolumn{1}{|c|}{\vo peak (mL/kg/min)} & \multicolumn{1}{c|}{32.12 (8.32) {[}15.90, 52.40{]}}   & \multicolumn{1}{c|}{27.32 (6.76) {[}15.90, 44.60{]}}   & 37.16 (6.69) {[}24.20, 52.40{]}   \\ \hline
\end{tabular}
}
\caption{Descriptive analysis of the clinical features and the target (\vo peak dataset).}
\end{table}

\begin{table}[ht!]
\centering
\resizebox{\columnwidth}{!}{
\begin{tabular}{c|ccc|}
\cline{2-4}
\multicolumn{1}{l|}{}                     & \multicolumn{3}{c|}{Group}                                                                                                                          \\ \hline
\multicolumn{1}{|c|}{Feature}            & \multicolumn{1}{c|}{All (n = 147)}                     & \multicolumn{1}{c|}{Women (n = 72)}                    & Men (n = 75)                      \\ \hline \hline
\multicolumn{1}{|c|}{Age (years)}         & \multicolumn{1}{c|}{22.30 (6.16) {[}11.33, 38.76{]}}   & \multicolumn{1}{c|}{22.48 (6.39) {[}11.33, 38.76{]}}   & 22.11 (5.95) {[}11.40, 36.92{]}   \\ \hline
\multicolumn{1}{|c|}{\acrshort{dt} (years)}          & \multicolumn{1}{c|}{2.13 (0.17) {[}1.17, 2.83{]}}      & \multicolumn{1}{c|}{2.15 (0.18) {[}1.17, 2.83{]}}      & 2.11 (0.15) {[}1.33, 2.41{]}      \\ \hline
\multicolumn{1}{|c|}{\acrshort{hrend} (bpm)} & \multicolumn{1}{l|}{150.65 (21.95) {[}75.00, 191.00{]}}  & \multicolumn{1}{l|}{145.97 (23.23) {[}75.00, 191.00{]}}  & \multicolumn{1}{l|}{155.47 (19.57) {[}99.00, 190.00{]}}  \\ \hline
\multicolumn{1}{|c|}{\acrshort{mvlpa} (min/day)}     & \multicolumn{1}{c|}{25.13 (25.81) {[}0.00, 149.18{]}}  & \multicolumn{1}{c|}{23.61 (26.59) {[}0.00, 149.18{]}}  & 26.70 (25.06) {[}0.00, 110.16{]}  \\ \hline
\multicolumn{1}{|c|}{Weight (kg)}         & \multicolumn{1}{c|}{67.19 (16.59) {[}40.50, 113.00{]}} & \multicolumn{1}{c|}{64.68 (17.39) {[}40.50, 113.00{]}} & 69.81 (15.39) {[}43.00, 108.00{]} \\ \hline
\multicolumn{1}{|c|}{\acrshort{6mwd} (m)}   & \multicolumn{1}{l|}{615.06 (77.21) {[}420.00, 873.00{]}} & \multicolumn{1}{l|}{639.97 (76.75) {[}420.00, 873.00{]}} & \multicolumn{1}{l|}{590.49 (69.88) {[}445.00, 755.00{]}} \\ \hline \hline
\multicolumn{1}{|c|}{\vo peak (mL/kg/min)} & \multicolumn{1}{c|}{32.12 (8.43) {[}15.90, 52.40{]}}   & \multicolumn{1}{c|}{27.19 (6.84) {[}15.90, 44.60{]}}   & 37.27 (6.69) {[}24.20, 52.40{]}   \\ \hline
\end{tabular}
}
\caption{Descriptive analysis of the clinical features and the target (\vo peak \textit{learning set}).}
\end{table}

\begin{table}[ht!]
\centering
\resizebox{\columnwidth}{!}{
\begin{tabular}{c|ccc|}
\cline{2-4}
\multicolumn{1}{l|}{}                     & \multicolumn{3}{c|}{Group}                                                                                                                    \\ \hline
\multicolumn{1}{|c|}{Feature}            & \multicolumn{1}{c|}{All (n = 17)}                    & \multicolumn{1}{c|}{Women (n = 8)}                   & Men (n = 9)                     \\ \hline \hline
\multicolumn{1}{|c|}{Age (years)}         & \multicolumn{1}{c|}{23.25 (5.98) {[}14.10, 34.43{]}} & \multicolumn{1}{c|}{22.72 (4.84) {[}14.10, 30.67{]}} & 23.85 (7.36) {[}15.46, 34.43{]} \\ \hline
\multicolumn{1}{|c|}{\acrshort{dt} (years)}          & \multicolumn{1}{c|}{2.16 (0.20) {[}1.59, 2.58{]}}    & \multicolumn{1}{c|}{2.23 (0.17) {[}2.00, 2.58{]}}    & 2.07 (0.22) {[}1.59, 2.25{]}    \\ \hline
\multicolumn{1}{|c|}{\acrshort{hrend} (bpm)} & \multicolumn{1}{l|}{138.94 (23.34) {[}96.00, 177.00{]}}  & \multicolumn{1}{l|}{135.25 (22.08) {[}112.00, 171.00{]}}  & \multicolumn{1}{l|}{142.22 (25.23) {[}96.00, 177.00{]}}  \\ \hline
\multicolumn{1}{|c|}{\acrshort{mvlpa} (min/day)}     & \multicolumn{1}{c|}{26.46 (17.29) {[}0.00, 50.49{]}} & \multicolumn{1}{c|}{28.11 (16.42) {[}0.22, 50.49{]}} & 24.61 (19.18) {[}0.00, 50.16{]} \\ \hline
\multicolumn{1}{|c|}{Weight (kg)}         & \multicolumn{1}{c|}{62.95 (7.73) {[}46.00, 73.60{]}} & \multicolumn{1}{c|}{61.80 (8.61) {[}46.00, 73.60{]}} & 64.25 (6.94) {[}53.00, 73.00{]} \\ \hline
\multicolumn{1}{|c|}{\acrshort{6mwd} (m)}     & \multicolumn{1}{l|}{588.94 (87.28) {[}412.00, 750.00{]}} & \multicolumn{1}{l|}{616.12 (112.84) {[}412.00, 750.00{]}} & \multicolumn{1}{l|}{564.78 (51.96) {[}480.00, 631.00{]}} \\ \hline \hline
\multicolumn{1}{|c|}{\vo peak (mL/kg/min)} & \multicolumn{1}{c|}{32.10 (7.60) {[}18.00, 46.80{]}} & \multicolumn{1}{c|}{28.44 (6.31) {[}18.00, 39.30{]}} & 36.21 (7.08) {[}27.00, 46.80{]} \\ \hline 
\end{tabular}
}
\caption{Descriptive analysis of the clinical features and the target (\vo peak \textit{holdout set}).}
\end{table}

\begin{table}[ht!]
\centering
\resizebox{\columnwidth}{!}{
\begin{tabular}{c|ccc|}
\cline{2-4}
\multicolumn{1}{c|}{}                          & \multicolumn{3}{c|}{Group}                                                                                                                                                     \\ \hline
\multicolumn{1}{|c|}{Feature}                 & \multicolumn{1}{c|}{All (n = 223)}                              & \multicolumn{1}{c|}{Women (n = 112)}                            & Men (n = 111)                              \\ \hline \hline
\multicolumn{1}{|c|}{Age at diagnosis (years)} & \multicolumn{1}{c|}{6.00 (4.54) {[}0.00, 18.00{]}}              & \multicolumn{1}{c|}{5.83 (4.41) {[}0.00, 17.00{]}}              & 6.17 (4.67) {[}1.00, 18.00{]}              \\ \hline
\multicolumn{1}{|c|}{\acrshort{bmi} at the \acrshort{eot} (kg/m$^{2}$)}       & \multicolumn{1}{c|}{18.55 (3.16) {[}12.71, 31.52{]}}            & \multicolumn{1}{c|}{18.89 (3.44) {[}12.71, 31.52{]}}            & 18.21 (2.82) {[}13.98, 26.43{]}            \\ \hline
\multicolumn{1}{|c|}{Doxorubicin (mg/m$^{2}$)}      & \multicolumn{1}{c|}{182.09 (120.22) {[}41.51, 472.85{]}}        & \multicolumn{1}{c|}{181.79 (123.45) {[}41.51, 472.85{]}}        & 182.40 (117.43) {[}43.64, 403.09{]}        \\ \hline
\multicolumn{1}{|c|}{\acrshort{dt} (years)}               & \multicolumn{1}{c|}{2.14 (0.17) {[}1.17, 2.83{]}}               & \multicolumn{1}{c|}{2.15 (0.19) {[}1.17, 2.83{]}}               & 2.12 (0.15) {[}1.33, 2.41{]}               \\ \hline
\multicolumn{1}{|c|}{\acrshort{eccd} (mg/m$^{2}$)}             & \multicolumn{1}{c|}{11088.88 (4992.14) {[}4028.75, 24918.16{]}} & \multicolumn{1}{c|}{10684.69 (4808.98) {[}4424.10, 24918.16{]}} & 11496.72 (5159.98) {[}4028.75, 24095.31{]} \\ \hline
\multicolumn{1}{|c|}{Methotrexate (mg/m$^{2}$)}     & \multicolumn{1}{c|}{6287.94 (1578.23) {[}853.63, 12784.10{]}}   & \multicolumn{1}{c|}{6113.30 (1539.55) {[}853.63, 9746.50{]}}    & 6464.15 (1603.98) {[}1036.32, 12784.10{]}  \\
\hline
\end{tabular}
}
\caption{Descriptive analysis of the numerical clinical features and the target (Obesity \textit{dataset}).}
\end{table}

\begin{table}[ht!]
\centering
\resizebox{0.75\columnwidth}{!}{
\begin{tabular}{cc|ccc|}
\cline{3-5}
\multicolumn{1}{l}{}                                                    & \multicolumn{1}{l|}{}         & \multicolumn{3}{c|}{Group}                                                                \\ \hline
\multicolumn{1}{|c|}{Feature}                                           & Modality                      & \multicolumn{1}{c|}{All (n = 223)} & \multicolumn{1}{c|}{Women (n = 112)} & Men (n = 111) \\ \hline \hline
\multicolumn{1}{|c|}{\multirow{3}{*}{Dexrazoxane (mg/m$^{2}$)}}              & 0                             & \multicolumn{1}{c|}{157 (70.40\%)} & \multicolumn{1}{c|}{79 (70.54\%)}    & 78 (70.27\%)  \\ \cline{2-5} 
\multicolumn{1}{|c|}{}                                                  & \textgreater{}0,$\leq$2963.16 & \multicolumn{1}{c|}{32 (14.35\%)}  & \multicolumn{1}{c|}{13 (11.61\%)}    & 19 (17.12\%)  \\ \cline{2-5} 
\multicolumn{1}{|c|}{}                                                  & \textgreater{}2963.16         & \multicolumn{1}{c|}{34 (15.25\%)}  & \multicolumn{1}{c|}{20 (17.86\%)}    & 14 (12.61\%)  \\ \hline
\multicolumn{1}{|c|}{\multirow{2}{*}{Gestational age at birth (weeks)}} & \textless{}37                 & \multicolumn{1}{c|}{13 (6.19\%)}   & \multicolumn{1}{c|}{10 (9.35\%)}     & 3 (2.91\%)    \\ \cline{2-5} 
\multicolumn{1}{|c|}{}                                                  & $\geq$37                      & \multicolumn{1}{c|}{197 (93.81\%)} & \multicolumn{1}{c|}{97 (90.65\%)}    & 100 (97.09\%) \\ \hline
\multicolumn{1}{|c|}{\multirow{2}{*}{Radiotherapy dose (Gy)}}           & 0                             & \multicolumn{1}{c|}{93 (41.70\%)}  & \multicolumn{1}{c|}{54 (48.21\%)}    & 39 (35.14\%)  \\ \cline{2-5} 
\multicolumn{1}{|c|}{}                                                  & \textgreater{}0               & \multicolumn{1}{c|}{130 (58.30\%)} & \multicolumn{1}{c|}{58 (51.79\%)}    & 72 (64.86\%)  \\ \hline
\end{tabular}
}
\caption{Descriptive analysis of the categorical clinical features (Obesity dataset).}
\end{table}

\begin{table}[ht!]
\centering
\resizebox{\columnwidth}{!}{
\begin{tabular}{c|ccc|}
\cline{2-4}
\multicolumn{1}{c|}{}                           & \multicolumn{3}{c|}{Group}                                                                                                                                                     \\ \hline
\multicolumn{1}{|c|}{Feature}                  & \multicolumn{1}{c|}{All (n = 200)}                              & \multicolumn{1}{c|}{Women (n = 100)}                            & Men (n = 100)                              \\ \hline \hline
\multicolumn{1}{|c|}{Age at diagnosis (years)}  & \multicolumn{1}{c|}{6.07 (4.56) {[}0.00, 18.00{]}}              & \multicolumn{1}{c|}{5.94 (4.47) {[}0.00, 17.00{]}}              & 6.20 (4.66) {[}1.00, 18.00{]}              \\ \hline
\multicolumn{1}{|c|}{\acrshort{bmi} at the \acrshort{eot} (kg/m$^{2}$)}   & \multicolumn{1}{c|}{18.53 (3.25) {[}12.71, 31.52{]}}            & \multicolumn{1}{c|}{18.86 (3.57) {[}12.71, 31.52{]}}            & 18.20 (2.87) {[}13.98, 26.43{]}            \\ \hline
\multicolumn{1}{|c|}{Doxorubicin (mg/m$^{2}$)}  & \multicolumn{1}{c|}{188.74 (120.12) {[}41.51, 472.85{]}}        & \multicolumn{1}{c|}{192.22 (123.18) {[}41.51, 472.85{]}}        & 185.27 (117.50) {[}43.64, 403.09{]}        \\ \hline
\multicolumn{1}{|c|}{\acrshort{dt} (years)}                & \multicolumn{1}{c|}{2.14 (0.18) {[}1.17, 2.83{]}}               & \multicolumn{1}{c|}{2.15 (0.20) {[}1.17, 2.83{]}}               & 2.12 (0.16) {[}1.33, 2.41{]}               \\ \hline
\multicolumn{1}{|c|}{\acrshort{eccd} (mg/m$^{2}$)}         & \multicolumn{1}{c|}{11251.40 (5069.78) {[}4028.75, 24918.16{]}} & \multicolumn{1}{c|}{10982.62 (4956.43) {[}4424.10, 24918.16{]}} & 11520.18 (5191.62) {[}4028.75, 24095.31{]} \\ \hline
\multicolumn{1}{|c|}{Methotrexate (mg/$m^{2}$)} & \multicolumn{1}{c|}{6286.33 (1575.20) {[}853.63, 12784.10{]}}   & \multicolumn{1}{c|}{6132.98 (1468.53) {[}853.63, 8035.45{]}}    & 6439.68 (1668.38) {[}1036.32, 12784.10{]}  \\
\hline
\end{tabular}
}
\caption{Descriptive analysis of the numerical clinical features and the target (Obesity \textit{learning set}).}
\end{table}

\begin{table}[ht!]
\centering
\resizebox{0.75\columnwidth}{!}{
\begin{tabular}{cc|ccc|}
\cline{3-5}
\multicolumn{1}{l}{}                                                    & \multicolumn{1}{l|}{}         & \multicolumn{3}{c|}{Group}                                                                \\ \hline
\multicolumn{1}{|c|}{Feature}                                           & Modality                      & \multicolumn{1}{c|}{All (n = 200)} & \multicolumn{1}{c|}{Women (n = 100)} & Men (n = 100) \\ \hline \hline
\multicolumn{1}{|c|}{\multirow{3}{*}{Dexrazoxane (mg/m$^2$)}}              & 0                             & \multicolumn{1}{c|}{140 (70.00\%)} & \multicolumn{1}{c|}{69 (69.00\%)}    & 71 (71.00\%)  \\ \cline{2-5} 
\multicolumn{1}{|c|}{}                                                  & \textgreater{}0,$\leq$2963.16 & \multicolumn{1}{c|}{29 (14.50\%)}  & \multicolumn{1}{c|}{12 (12.00\%)}    & 17 (17.00\%)  \\ \cline{2-5} 
\multicolumn{1}{|c|}{}                                                  & \textgreater{}2963.16         & \multicolumn{1}{c|}{31 (15.50\%)}  & \multicolumn{1}{c|}{19 (19.00\%)}    & 12 (12.00\%)  \\ \hline
\multicolumn{1}{|c|}{\multirow{2}{*}{Gestational age at birth (weeks)}} & \textless{}37                 & \multicolumn{1}{c|}{11 (5.85\%)}   & \multicolumn{1}{c|}{9 (9.47\%)}      & 2 (2.15\%)    \\ \cline{2-5} 
\multicolumn{1}{|c|}{}                                                  & $\geq$37                      & \multicolumn{1}{c|}{177 (94.15\%)} & \multicolumn{1}{c|}{86 (90.53\%)}    & 91 (97.85\%)  \\ \hline
\multicolumn{1}{|c|}{\multirow{2}{*}{Radiotherapy dose (Gy)}}           & 0                             & \multicolumn{1}{c|}{79 (39.50\%)}  & \multicolumn{1}{c|}{43 (43.00\%)}    & 36 (36.00\%)  \\ \cline{2-5} 
\multicolumn{1}{|c|}{}                                                  & \textgreater{}0               & \multicolumn{1}{c|}{121 (60.50\%)} & \multicolumn{1}{c|}{57 (57.00\%)}    & 64 (64.00\%)  \\ \hline
\end{tabular}
}
\caption{Descriptive analysis of the categorical clinical features (Obesity \textit{learning set}).}
\end{table}

\begin{table}[ht!]
\centering
\resizebox{\columnwidth}{!}{
\begin{tabular}{c|ccc|}
\cline{2-4}
\multicolumn{1}{c|}{}                           & \multicolumn{3}{c|}{Group}                                                                                                                                                   \\ \hline
\multicolumn{1}{|c|}{Feature}                  & \multicolumn{1}{c|}{All (n = 23)}                              & \multicolumn{1}{c|}{Women (n = 12)}                            & Men (n = 11)                               \\ \hline \hline
\multicolumn{1}{|c|}{Age at diagnosis (years)}  & \multicolumn{1}{c|}{5.39 (4.43) {[}2.00, 16.00{]}}             & \multicolumn{1}{c|}{4.92 (3.96) {[}2.00, 16.00{]}}             & 5.91 (5.03) {[}2.00, 16.00{]}              \\ \hline
\multicolumn{1}{|c|}{\acrshort{bmi} at the \acrshort{eot} (kg/m$^{2}$)}   & \multicolumn{1}{c|}{18.72 (2.28) {[}14.55, 23.01{]}}           & \multicolumn{1}{c|}{19.18 (2.10) {[}16.47, 23.01{]}}           & 18.30 (2.45) {[}14.55, 22.12{]}            \\ \hline
\multicolumn{1}{|c|}{Doxorubicin (mg/m$^{2}$)}  & \multicolumn{1}{c|}{124.27 (106.89) {[}44.23, 301.39{]}}       & \multicolumn{1}{c|}{94.87 (89.37) {[}44.23, 299.34{]}}         & 156.35 (119.08) {[}46.43, 301.39{]}        \\ \hline
\multicolumn{1}{|c|}{\acrshort{dt} (years)}                & \multicolumn{1}{c|}{2.15 (0.09) {[}2.00, 2.25{]}}              & \multicolumn{1}{c|}{2.17 (0.09) {[}2.00, 2.25{]}}              & 2.13 (0.09) {[}2.00, 2.25{]}               \\ \hline
\multicolumn{1}{|c|}{\acrshort{eccd} (mg/m$^{2}$)}         & \multicolumn{1}{c|}{9675.70 (4081.64) {[}6049.08, 21437.53{]}} & \multicolumn{1}{c|}{8201.96 (2176.78) {[}6049.08, 13413.93{]}} & 11283.41 (5098.02) {[}6636.30, 21437.53{]} \\ \hline
\multicolumn{1}{|c|}{Methotrexate (mg/$m^{2}$)} & \multicolumn{1}{c|}{6301.94 (1640.15) {[}2104.66, 9746.50{]}}  & \multicolumn{1}{c|}{5949.30 (2115.48) {[}2104.66, 9746.50{]}}  & 6686.64 (826.67) {[}4986.33, 7687.01{]}    \\
\hline
\end{tabular}
}
\caption{Descriptive analysis of the numerical clinical features and the target (Obesity \textit{holdout set}).}
\end{table}

\begin{table}[ht!]
\centering
\resizebox{0.75\columnwidth}{!}{
\begin{tabular}{cc|ccc|}
\cline{3-5}
\multicolumn{1}{l}{}                                                    & \multicolumn{1}{l|}{}         & \multicolumn{3}{c|}{Group}                                                                \\ \hline
\multicolumn{1}{|c|}{Feature}                                           & Modality                      & \multicolumn{1}{c|}{All (n = 23)} & \multicolumn{1}{c|}{Women (n = 12)} & Men (n = 11) \\ \hline \hline
\multicolumn{1}{|c|}{\multirow{3}{*}{Dexrazoxane (mg/m$^2$)}}              & 0                             & \multicolumn{1}{c|}{17 (73.91\%)}  & \multicolumn{1}{c|}{10 (83.33\%)}    & 7 (63.64\%)   \\ \cline{2-5} 
\multicolumn{1}{|c|}{}                                                  & \textgreater{}0,$\leq$2963.16 & \multicolumn{1}{c|}{3 (13.04\%)}   & \multicolumn{1}{c|}{1 (8.33\%)}      & 2 (18.18\%)   \\ \cline{2-5} 
\multicolumn{1}{|c|}{}                                                  & \textgreater{}2963.16         & \multicolumn{1}{c|}{3 (13.04\%)}   & \multicolumn{1}{c|}{1 (8.33\%)}      & 2 (18.18\%)   \\ \hline
\multicolumn{1}{|c|}{\multirow{2}{*}{Gestational age at birth (weeks)}} & \textless{}37                 & \multicolumn{1}{c|}{2 (9.09\%)}    & \multicolumn{1}{c|}{1 (8.33\%)}      & 1 (10.00\%)   \\ \cline{2-5} 
\multicolumn{1}{|c|}{}                                                  & $\geq$37                      & \multicolumn{1}{c|}{20 (90.91\%)}  & \multicolumn{1}{c|}{11 (91.67\%)}    & 9 (90.00\%)   \\ \hline
\multicolumn{1}{|c|}{\multirow{2}{*}{Radiotherapy dose (Gy)}}           & 0                             & \multicolumn{1}{c|}{14 (60.87\%)}  & \multicolumn{1}{c|}{11 (91.67\%)}    & 3 (27.27\%)   \\ \cline{2-5} 
\multicolumn{1}{|c|}{}                                                  & \textgreater{}0               & \multicolumn{1}{c|}{9 (39.13\%)}   & \multicolumn{1}{c|}{1 (8.33\%)}      & 8 (72.73\%)   \\ \hline
\end{tabular}
}
\caption{Descriptive analysis of the categorical clinical features (Obesity \textit{holdout set}).}
\end{table}

\clearpage

\begin{table}[ht!]
\centering
\resizebox{0.25\paperheight}{!}{
\begin{tabular}{cc|ccc|}
\cline{3-5}
                                                   &          & \multicolumn{3}{c|}{Group}                                                              \\ \hline
\multicolumn{1}{|c|}{\acrshort{snp}}                          & Modality & \multicolumn{1}{c|}{All (n = 223)}           & \multicolumn{1}{c|}{Women (n = 112)}         & Men (n = 111)         \\ \hline \hline
\multicolumn{1}{|c|}{\multirow{3}{*}{1:66036441}}  & 0/0      & \multicolumn{1}{c|}{123 (55.16\%)} & \multicolumn{1}{c|}{62 (55.36\%)}  & 61 (54.95\%)  \\ \cline{2-5} 
\multicolumn{1}{|c|}{}                             & 0/1      & \multicolumn{1}{c|}{82 (36.77\%)}  & \multicolumn{1}{c|}{43 (38.39\%)}  & 39 (35.14\%)  \\ \cline{2-5} 
\multicolumn{1}{|c|}{}                             & 1/1      & \multicolumn{1}{c|}{18 (8.07\%)}   & \multicolumn{1}{c|}{7 (6.25\%)}    & 11 (9.91\%)   \\ \hline
\multicolumn{1}{|c|}{\multirow{2}{*}{1:161182208}} & 0/0      & \multicolumn{1}{c|}{179 (80.63\%)} & \multicolumn{1}{c|}{88 (79.28\%)}  & 91 (81.98\%)  \\ \cline{2-5} 
\multicolumn{1}{|c|}{}                             & 0/1      & \multicolumn{1}{c|}{43 (19.37\%)}  & \multicolumn{1}{c|}{23 (20.72\%)}  & 20 (18.02\%)  \\ \hline
\multicolumn{1}{|c|}{\multirow{3}{*}{1:226019633}} & 0/0      & \multicolumn{1}{c|}{110 (49.33\%)} & \multicolumn{1}{c|}{56 (50.00\%)}  & 54 (48.65\%)  \\
\hline
\end{tabular}
}
\caption{Descriptive analysis of the \acrshort{snp}s (Obesity dataset).}
\end{table}

\newpage 

\begin{table}[ht!]
\centering
\resizebox{0.25\paperheight}{!}{
\begin{tabular}{cc|lll|}
\cline{3-5}
                                                   &          & \multicolumn{3}{c|}{Group}                                                                        \\ \hline
\multicolumn{1}{|c|}{\acrshort{snp}}                          & Modality & \multicolumn{1}{c|}{All (n = 200)}           & \multicolumn{1}{c|}{Women (n = 100)}        & \multicolumn{1}{c|}{Men (n = 100)} \\ \hline
\multicolumn{1}{|c|}{\multirow{3}{*}{1:66036441}}  & 0/0      & \multicolumn{1}{c|}{112 (56.00\%)} & \multicolumn{1}{c|}{59 (59.00\%)} & 53 (53.00\%)             \\ \cline{2-5} 
\multicolumn{1}{|c|}{}                             & 0/1      & \multicolumn{1}{c|}{72 (36.00\%)}  & \multicolumn{1}{c|}{35 (35.00\%)} & 37 (37.00\%)             \\ \cline{2-5} 
\multicolumn{1}{|c|}{}                             & 1/1      & \multicolumn{1}{c|}{16 (8.00\%)}   & \multicolumn{1}{c|}{6 (6.00\%)}   & 10 (10.00\%)             \\ \hline
\multicolumn{1}{|c|}{\multirow{2}{*}{1:161182208}} & 0/0      & \multicolumn{1}{c|}{160 (80.40\%)} & \multicolumn{1}{c|}{77 (77.78\%)} & 83 (83.00\%)             \\ \cline{2-5} 
\multicolumn{1}{|c|}{}                             & 0/1      & \multicolumn{1}{c|}{39 (19.60\%)}  & \multicolumn{1}{c|}{22 (22.22\%)} & 17 (17.00\%)             \\ \hline
\multicolumn{1}{|c|}{\multirow{3}{*}{1:226019633}} & 0/0      & \multicolumn{1}{c|}{97 (48.50\%)}  & \multicolumn{1}{c|}{49 (49.00\%)} & 48 (48.00\%)             \\
\hline
\end{tabular}
}
\caption{Descriptive analysis of the \acrshort{snp}s (Obesity \textit{learning set}).}
\end{table}

\newpage

\begin{table}[ht!]
\centering
\resizebox{0.25\paperheight}{!}{
\begin{tabular}{cc|ccc|}
\cline{3-5}
                                                   &          & \multicolumn{3}{c|}{Group}                                                            \\ \hline
\multicolumn{1}{|c|}{\acrshort{snp}}                          & Modality & \multicolumn{1}{c|}{All (n = 23)}          & \multicolumn{1}{c|}{Women (n = 12)}         & \multicolumn{1}{c|}{Men (n = 11)}          \\ \hline \hline
\multicolumn{1}{|c|}{\multirow{3}{*}{1:66036441}}  & 0/0      & \multicolumn{1}{c|}{11 (47.83\%)} & \multicolumn{1}{c|}{3 (25.00\%)}   & 8 (72.73\%)  \\ \cline{2-5} 
\multicolumn{1}{|c|}{}                             & 0/1      & \multicolumn{1}{c|}{10 (43.48\%)} & \multicolumn{1}{c|}{8 (66.67\%)}   & 2 (18.18\%)  \\ \cline{2-5} 
\multicolumn{1}{|c|}{}                             & 1/1      & \multicolumn{1}{c|}{2 (8.70\%)}   & \multicolumn{1}{c|}{1 (8.33\%)}    & 1 (9.09\%)   \\ \hline
\multicolumn{1}{|c|}{\multirow{2}{*}{1:161182208}} & 0/0      & \multicolumn{1}{c|}{19 (82.61\%)} & \multicolumn{1}{c|}{11 (91.67\%)}  & 8 (72.73\%)  \\ \cline{2-5} 
\multicolumn{1}{|c|}{}                             & 0/1      & \multicolumn{1}{c|}{4 (17.39\%)}  & \multicolumn{1}{c|}{1 (8.33\%)}    & 3 (27.27\%)  \\ \hline
\multicolumn{1}{|c|}{\multirow{3}{*}{1:226019633}} & 0/0      & \multicolumn{1}{c|}{13 (56.52\%)} & \multicolumn{1}{c|}{7 (58.33\%)}   & 6 (54.55\%)  \\
\hline
\end{tabular}
}
\caption{Descriptive analysis of the \acrshort{snp}s (Obesity \textit{holdout set}).}
\end{table}

\newpage

\subsection{Evaluation of the models}
In this subsection, we present tables with the scores obtained by the models following the 10-splits random stratified subsampling evaluation on the \textit{learning set} associated to each prediction task. The scores are reported into the format \textit{mean} $\pm$ \textit{\acrshort{std}}.

\begin{table}[ht!]
\centering
\resizebox{0.78\columnwidth}{!}{
\begin{tabular}{cc|cccc|}
\cline{3-6}
\multicolumn{1}{l}{}                             & \multicolumn{1}{l|}{}      & \multicolumn{4}{c|}{Metric}                                                                             \\ \hline
\multicolumn{1}{|c|}{\acrshort{hp}s optimization}            & Model                      & \multicolumn{1}{c|}{\acrshort{rmse}} & \multicolumn{1}{c|}{\acrshort{mae}} & \multicolumn{1}{c|}{\acrshort{pcc}} & \multicolumn{1}{c|}{\acrshort{cindex}}              \\ \hline \hline
\multicolumn{1}{|c|}{\multirow{6}{*}{Manual}}    & \textbf{Random forest}     & 5.44 $\pm$ 0.74              & 4.28 $\pm$ 0.54             & 0.79 $\pm$ 0.05             & 0.81 $\pm$ 0.02          \\
\multicolumn{1}{|c|}{}                           & \textbf{XGBoost}           & 5.64 $\pm$ 0.90               & 4.34 $\pm$ 0.70              & 0.76 $\pm$ 0.08             & 0.79 $\pm$ 0.04          \\
\multicolumn{1}{|c|}{}                           & \textbf{Linear regression} & 5.45 $\pm$ 0.71              & 4.27 $\pm$ 0.58             & 0.78 $\pm$ 0.07             & 0.80 $\pm$ 0.04           \\
\multicolumn{1}{|c|}{}                           & \textbf{\acrshort{mlp}}               & 6.00 $\pm$ 1.05               & 4.54 $\pm$ 0.75             & 0.78 $\pm$ 0.08             & 0.80 $\pm$ 0.04           \\
\multicolumn{1}{|c|}{}                           & \textbf{\acrshort{gcn}}               & 5.42 $\pm$ 0.86              & 4.14 $\pm$ 0.60              & 0.79 $\pm$ 0.07             & 0.81 $\pm$ 0.03          \\
\multicolumn{1}{|c|}{}                           & \textbf{\acrshort{gat}}               & \textbf{5.34 $\pm$ 0.80}      & 4.13 $\pm$ 0.59             & \textbf{0.80 $\pm$ 0.07}     & 0.81 $\pm$ 0.03          \\
\hline
\end{tabular}
}
\caption[Evaluation of the models during the \vo peak prediction task.]{Scores obtained by the models following the 10-splits random stratified subsampling evaluation on the \textit{learning set} of the \vo peak prediction task.}
\end{table}

\begin{table}[ht!]
\centering
\resizebox{\columnwidth}{!}{
\begin{tabular}{cc|cccccc|}
\cline{3-8}
\multicolumn{1}{l}{}                             & \multicolumn{1}{l|}{}      & \multicolumn{6}{c|}{Metric}                                                                                                                                                  \\ \hline
\multicolumn{1}{|c|}{\acrshort{hp}s optimization}            & Model                      & \multicolumn{1}{c|}{\acrshort{rmse}} & \multicolumn{1}{c|}{\acrshort{mae}} & \multicolumn{1}{c|}{\acrshort{pcc}} & \multicolumn{1}{c|}{\acrshort{cindex}}     & \multicolumn{1}{c|}{Sensitivity} & \multicolumn{1}{c|}{Specificity}         \\ \hline \hline
\multicolumn{1}{|c|}{\multirow{6}{*}{Manual}}    & \textbf{Random forest}     & 8.97 $\pm$ 0.53              & 7.36 $\pm$ 0.46             & 0.65 $\pm$ 0.04             & 0.73 $\pm$ 0.02                     & 0.80 $\pm$ 0.06                      & 0.71 $\pm$ 0.12         \\
\multicolumn{1}{|c|}{}                           & \textbf{XGBoost}           & 9.12 $\pm$ 0.44              & 7.31 $\pm$ 0.35             & 0.65 $\pm$ 0.05             & 0.73 $\pm$ 0.02                     & 0.79 $\pm$ 0.05                     & 0.74 $\pm$ 0.09         \\
\multicolumn{1}{|c|}{}                           & \textbf{Linear regression} & 8.96 $\pm$ 0.62              & 7.39 $\pm$ 0.59             & \textbf{0.66 $\pm$ 0.05}    & \textbf{0.75 $\pm$ 0.02}            & 0.75 $\pm$ 0.08                     & 0.72 $\pm$ 0.12         \\
\multicolumn{1}{|c|}{}                           & \textbf{\acrshort{mlp}}               & 9.07 $\pm$ 0.59              & 7.27 $\pm$ 0.48             & 0.64 $\pm$ 0.07             & 0.74 $\pm$ 0.04                     & 0.68 $\pm$ 0.10                     & \textbf{0.76 $\pm$ 0.10} \\
\multicolumn{1}{|c|}{}                           & \textbf{\acrshort{gcn}}                & 9.03 $\pm$ 0.61     & 7.47 $\pm$ 0.53  & 0.65 $\pm$ 0.04    & 0.74 $\pm$ 0.02       & 0.76 $\pm$ 0.12            & 0.75 $\pm$ 0.10 \\
\hline
\end{tabular}
}
\caption[Evaluation of the models (w/o \acrshort{snp}s) during the obesity prediction task.]{Scores obtained by the models following the 10-splits random stratified subsampling evaluation without \acrshort{snp}s on the \textit{learning set} of the obesity prediction task.}
\end{table}

\begin{table}[ht!]
\centering
\resizebox{\columnwidth}{!}{
\begin{tabular}{cc|cccccc|}
\cline{3-8}
                                                 &                            & \multicolumn{6}{c|}{Metric}                                                                                                                                               \\ \hline
\multicolumn{1}{|c|}{\acrshort{hp}s optimization}            & Model                      & \multicolumn{1}{c|}{\acrshort{rmse}} & \multicolumn{1}{c|}{\acrshort{mae}} & \multicolumn{1}{c|}{\acrshort{pcc}} & \multicolumn{1}{c|}{\acrshort{cindex}} & \multicolumn{1}{c|}{Sensitivity} & Specificity           \\ \hline \hline
\multicolumn{1}{|c|}{\multirow{7}{*}{Manual}}    & \textbf{Random forest}     & 9.64 $\pm$ 0.71              & 7.84 $\pm$ 0.67             & 0.62 $\pm$ 0.04             & 0.73 $\pm$ 0.01                 & \textbf{0.79 $\pm$ 0.09}            & 0.67 $\pm$ 0.09          \\
\multicolumn{1}{|c|}{}                           & \textbf{XGBoost}           & 9.53 $\pm$ 0.53              & 7.80 $\pm$ 0.34              & 0.60 $\pm$ 0.04              & 0.71 $\pm$ 0.02                 & 0.74 $\pm$ 0.08                     & 0.70 $\pm$ 0.10            \\
\multicolumn{1}{|c|}{}                           & \textbf{Linear regression} & 10.16 $\pm$ 0.73             & 8.35 $\pm$ 0.69             & 0.52 $\pm$ 0.08             & 0.67 $\pm$ 0.03                 & 0.62 $\pm$ 0.13                     & 0.70 $\pm$ 0.11            \\
\multicolumn{1}{|c|}{}                           & \textbf{\acrshort{mlp}}               & 11.63 $\pm$ 1.59             & 9.05 $\pm$ 1.03             & 0.44 $\pm$ 0.12             & 0.66 $\pm$ 0.03                 & 0.59 $\pm$ 0.10                     &    0.73 $\pm$ 0.11          \\
\multicolumn{1}{|c|}{}                           & \textbf{\acrshort{gcn}}               & 11.46 $\pm$ 1.03             & 9.26 $\pm$ 0.64             & 0.46 $\pm$ 0.10              & 0.65 $\pm$ 0.04                 & 0.62 $\pm$ 0.14                      & 0.69 $\pm$ 0.11           \\
\hline
\end{tabular}
}
\caption[Evaluation of the models (w/ \acrshort{snp}s) during the obesity prediction task.]{Scores obtained by the models following the 10-splits random stratified subsampling evaluation with \acrshort{snp}s on the \textit{learning set} of the obesity prediction task.}
\end{table}

\subsubsection{Evaluations of graphs' neighborhood sizes}
Here, we display tables that motivated the choices of neighborhood sizes used in the construction of the graphs associated to the GNNs.

\begin{table}[ht]
\centering
\begin{tabular}{cc|cccc|}
\cline{3-6}
                                           &                           & \multicolumn{4}{c|}{Metric}                                                                             \\ \hline
\multicolumn{1}{|c|}{Model}                & Number of neighbors  & \multicolumn{1}{c|}{RMSE} & \multicolumn{1}{c|}{MAE} & \multicolumn{1}{c|}{PCC} & C-Index               \\ \hline \hline
\multicolumn{1}{|c|}{\multirow{4}{*}{\textbf{GCN}}} & 4                         & 5.52 $\pm$ 0.81              & 4.17 $\pm$ 0.56             & \textbf{0.79 $\pm$ 0.07}    & 0.80 $\pm$ 0.04          \\
\multicolumn{1}{|c|}{}                     & 6                         & 5.46 $\pm$ 0.83              & 4.17 $\pm$ 0.59             & \textbf{0.79 $\pm$ 0.07}    & \textbf{0.81 $\pm$ 0.03} \\
\multicolumn{1}{|c|}{}                     & 8                         & \textbf{5.42 $\pm$ 0.86}     & \textbf{4.14 $\pm$ 0.60}    & \textbf{0.79 $\pm$ 0.07}    & \textbf{0.81 $\pm$ 0.03} \\
\multicolumn{1}{|c|}{}                     & 10                        & 5.45 $\pm$ 0.84              & 4.18 $\pm$ 0.57             & \textbf{0.79 $\pm$ 0.07}    & 0.81 $\pm$ 0.04          \\ \hline \hline
\multicolumn{1}{|c|}{\multirow{4}{*}{\textbf{GAT}}} & 4                         & 5.46 $\pm$ 0.86              & 4.19 $\pm$ 0.63             & 0.79 $\pm$ 0.06             & \textbf{0.81 $\pm$ 0.03} \\
\multicolumn{1}{|c|}{}                     & 6                         & 5.45 $\pm$ 0.82              & 4.21 $\pm$ 0.61             & 0.79 $\pm$ 0.06             & \textbf{0.81 $\pm$ 0.03} \\
\hline
\end{tabular}
\caption[Graph neighborhood evaluation (\vo peak)]{Graph neighborhood evaluation (\vo peak). Values of neighborhood size were tested with the other manually selected hyperparameters reported in Supplementary Tables S24-S25.}
\end{table}

\begin{table}[ht!]
\resizebox{\columnwidth}{!}{
\begin{tabular}{cc|cccccc|}
\cline{3-8}
                                           &                           & \multicolumn{6}{c|}{Metric}                                                                                                                                               \\ \hline
\multicolumn{1}{|c|}{Model}                & Number of neighbors  & \multicolumn{1}{c|}{RMSE} & \multicolumn{1}{c|}{MAE} & \multicolumn{1}{c|}{PCC} & \multicolumn{1}{c|}{C-Index} & \multicolumn{1}{c|}{Sensitivity} & Specificity           \\ \hline \hline
\multicolumn{1}{|c|}{\multirow{4}{*}{\textbf{GCN}}} & 4                         & 9.86 $\pm$ 0.70              & 7.99 $\pm$ 0.63             & 0.58 $\pm$ 0.06             & 0.70 $\pm$ 0.03                 & 0.67 $\pm$ 0.07                     & 0.71 $\pm$ 0.10          \\
\multicolumn{1}{|c|}{}                     & 6                         & 9.06 $\pm$ 0.60              & 7.47 $\pm$ 0.54             & 0.64 $\pm$ 0.05             & \textbf{0.74 $\pm$ 0.02}        & 0.75 $\pm$ 0.11                     & 0.74 $\pm$ 0.10          \\
\multicolumn{1}{|c|}{}                     & 8                         & \textbf{9.03 $\pm$ 0.61}     & \textbf{7.47 $\pm$ 0.53}    & \textbf{0.65 $\pm$ 0.04}    & \textbf{0.74 $\pm$ 0.02}        & \textbf{0.76 $\pm$ 0.12}            & \textbf{0.75 $\pm$ 0.10} \\
\multicolumn{1}{|c|}{}                     & 10                        & 9.84 $\pm$ 0.52              & 8.01 $\pm$ 0.52             & 0.58 $\pm$ 0.07             & 0.71 $\pm$ 0.03                 & 0.63 $\pm$ 0.11                     & 0.74 $\pm$ 0.12          \\ \hline \hline
\multicolumn{1}{|c|}{\multirow{4}{*}{\textbf{GAT}}} & 4                         & 9.85 $\pm$ 0.84              & 7.85 $\pm$ 0.74             & 0.57 $\pm$ 0.07             & 0.70 $\pm$ 0.03                 & 0.60 $\pm$ 0.14                     & \textbf{0.76 $\pm$ 0.09} \\
\hline 
\end{tabular}
}
\caption[Graph neighborhood evaluation (obesity w/o SNPs)]{Graph neighborhood evaluation (obesity w/o SNPs). Values of neighborhood size were tested with the other manually selected hyperparameters reported in Supplementary Tables S24-S25.}
\end{table}

\begin{table}[ht!]
\centering
\resizebox{\columnwidth}{!}{
\begin{tabular}{cc|cccccc|}
\cline{3-8}
                                           &                           & \multicolumn{6}{c|}{Metric}                                                                                                                                               \\ \hline
\multicolumn{1}{|c|}{Model}                & Number of neighbors  & \multicolumn{1}{c|}{RMSE} & \multicolumn{1}{c|}{MAE} & \multicolumn{1}{c|}{PCC} & \multicolumn{1}{c|}{C-Index} & \multicolumn{1}{c|}{Sensitivity} & Specificity           \\ \hline \hline
\multicolumn{1}{|c|}{\multirow{4}{*}{\textbf{GCN}}} & 4                         & 12.0 $\pm$ 1.48              & 9.62 $\pm$ 1.11             & 0.41 $\pm$ 0.11             & 0.64 $\pm$ 0.03                 & 0.59 $\pm$ 0.10                     & \textbf{0.78 $\pm$ 0.08} \\
\multicolumn{1}{|c|}{}                     & 6                         & 11.89 $\pm$ 0.99             & 9.46 $\pm$ 0.75             & 0.42 $\pm$ 0.09             & 0.64 $\pm$ 0.04                 & \textbf{0.64 $\pm$ 0.12}             & 0.71 $\pm$ 0.15          \\
\multicolumn{1}{|c|}{}                     & 8                         & \textbf{11.46 $\pm$ 1.03}    & \textbf{9.26 $\pm$ 0.64}    & \textbf{0.46 $\pm$ 0.10}     & 0.65 $\pm$ 0.04                 & 0.62 $\pm$ 0.14                      & 0.69 $\pm$ 0.11           \\
\multicolumn{1}{|c|}{}                     & 10                        & 12.15 $\pm$ 1.42             & 9.85 $\pm$ 1.38             & 0.43 $\pm$ 0.06             & \textbf{0.65 $\pm$ 0.02}        & 0.63 $\pm$ 0.12                     & 0.68 $\pm$ 0.13           \\ \hline \hline
\multicolumn{1}{|c|}{\multirow{4}{*}{\textbf{GAT}}} & 4                         & 11.12 $\pm$ 0.90              & 9.07 $\pm$ 0.81             & 0.40 $\pm$ 0.09              & 0.64 $\pm$ 0.03                 & 0.60 $\pm$ 0.14                      & 0.71 $\pm$ 0.12          \\
\hline
\end{tabular}
}
\caption[Graph neighborhood evaluation (obesity w/ SNPs)]{Graph neighborhood evaluation (obesity w/ SNPs). Values of neighborhood size were tested with the other manually selected hyperparameters reported in Supplementary Tables S24-S25.}
\end{table}

\newpage

\section{SUPPLEMENTARY METHODS}
\subsection{Construction of the datasets}
Figures reported in this subsection illustrate the construction procedures of the two datasets used in the study.
\begin{figure}[ht!]
    \centering
    \includegraphics[scale=0.72]{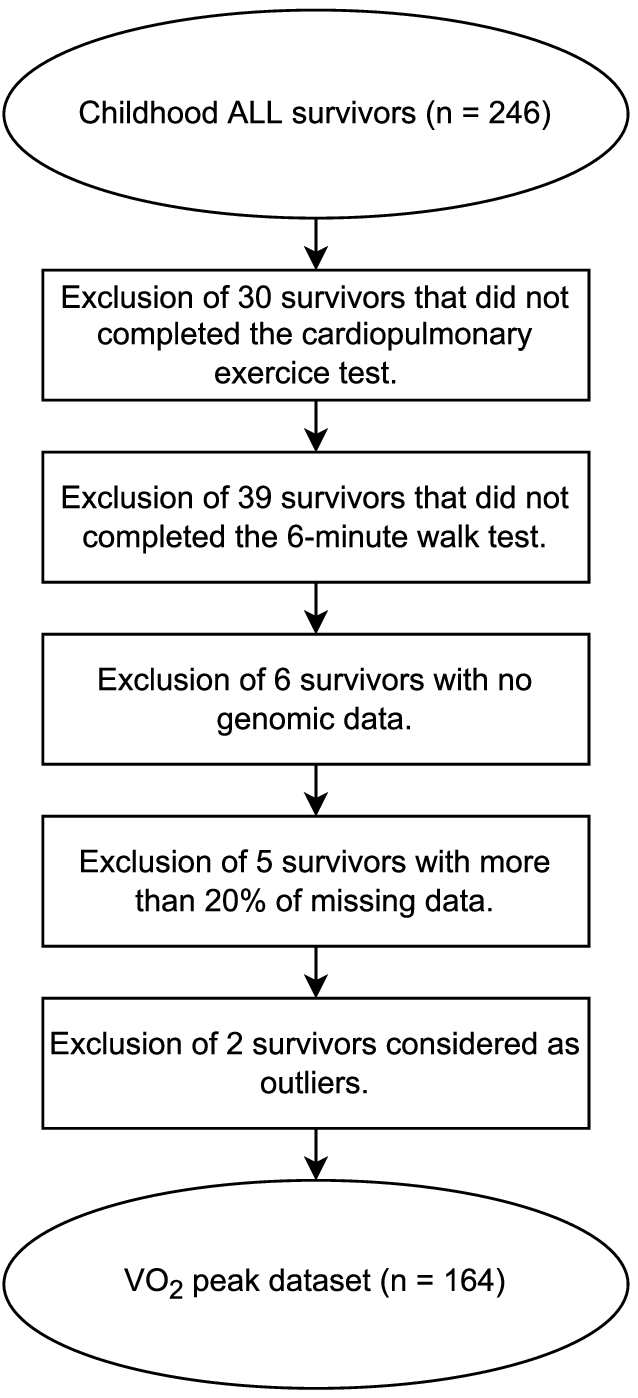}
    \caption[Construction of the \vo peak dataset.]{Construction of the \vo peak dataset. The 6 survivors without genomic data were excluded to allow the possibility of an eventual experiment including SNPs. The first outlier was excluded due to a short DT of 6 months. The second outlier was excluded considering its high MVLPA of 238 minutes per day.}
    \label{fig:vo2_dataset}
\end{figure}

\begin{figure}[ht!]
    \centering
    \includegraphics[scale=0.72]{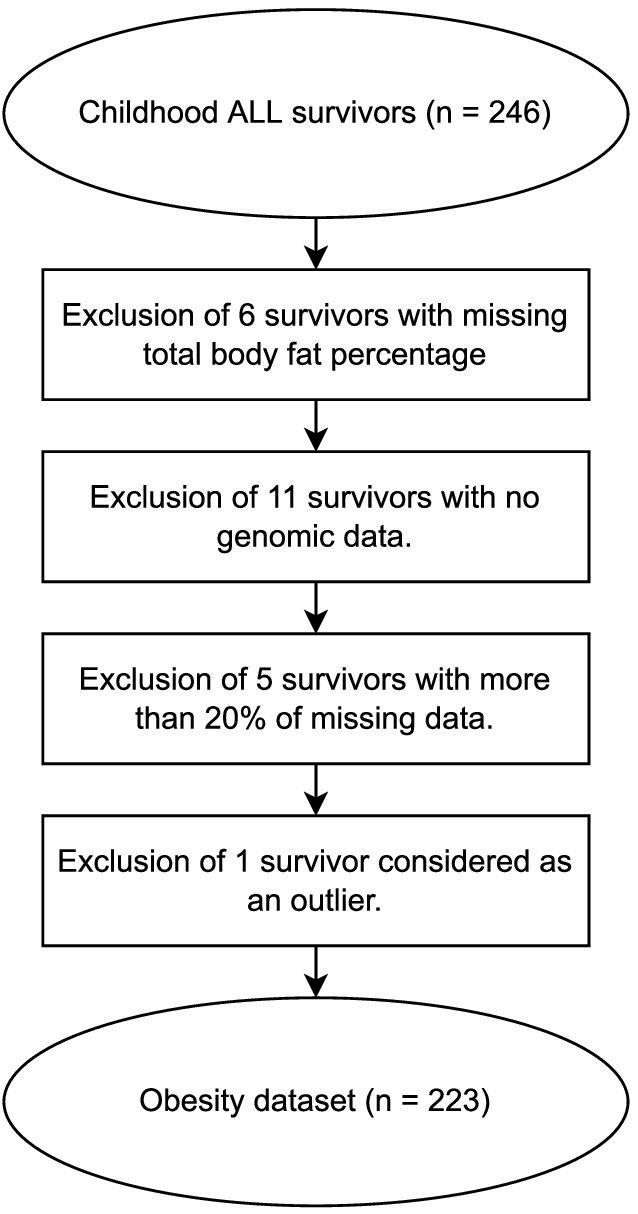}
    \caption[Construction of the obesity dataset.]{Construction of the obesity dataset. The outlier was excluded due to a short DT of 6 months.}
    \label{fig:obesity_dataset}
\end{figure}
\subsection{Architectures of the models}

In this subsection, we describe the model architectures that were implemented using \textit{Pytorch} \cite{Paszke:2019} and \textit{DGL} \cite{Wang:2019} librairies. We denote $\num$ and $\cat$ the sets of numerical and categorical features observed in a dataset and $k_j$ the number of modalities associated to any categorical feature $j \in \cat$. We use the notation $\xpoint \in \real^{(|\num| + |\cat|) \times 1}$ to represent a single data point in a dataset and $\xcoord$ its component associated to the feature $j \in \num \, \cup \, \cat$. We also identify the prediction associated to a datapoint $\xpoint$ as $y_{i}$. Hence, considering this notation, we can define $$\xpoint = \left[\xnum;\xcat\right]$$  where $\xnum = \left[\xcoord \right]_{j \in \num} \in \real^{|\num| \times 1}$ and $\xcat = \left[\xcoord \right]_{j \in \cat} \in \real^{|\cat| \times 1}$ contain the standardized numerical features and the nominal encodings of the categorical features respectively.

\subsubsection{Categorical embedding}
We integrated the categorical features within each of our architecture using categorical embeddings. More precisely, each categorical feature ($\xcoord$ s.t $j \in \cat$) is first mapped to a vectorial representation (i.e., embedding) $\emb \in \real^{k_j \times k_j}$. Each embedding is defined as $$\emb = \Wemb{j}\matij{o}{i}{j}$$ where $\Wemb{j} \in \real^{k_j \times k_j}$ is a matrix of parameters initialized independently from a standard normal distribution and $\matij{o}{i}{j} \in \real^{k_j \times 1}$ is a one-hot vector with the non-null value at the position represented by the nominal encoding $\xcoord$. \\

\noindent All categorical embeddings are further concatenated to represented the enriched representation of $\xcat$ that we denote $\xcatprime \in \real^{\sum_{j \in \cat}k_j \times 1}$.
$$\xcatprime = \Vert_{j\in \cat} \, \emb$$

\subsubsection{Linear regression}
The linear regression model was implemented such that
$$y_i = \w^{T}\left[\xnum;\xcatprime\right] + b$$
where $\w \in \real^{(|\num| + \sum_{j \in \cat}k_j) \times 1}$ and $b \in \real$.

\subsubsection{Multi-layer Perceptron (MLP)}
The \acrshort{mlp} was designed to have a single hidden layer containing half the number of features of the input layer. Its architecture is defined in order to have
$$y_i = \w^{T}\left(\sigma\left(\W\left[\xnum;\xcatprime\right] + \mati{b}{1}\right)\right) + b_2$$
where $\W \in \real^{round(n/2) \times n}$, $\w \in \real^{round(n/2) \times 1}$, $\mati{b}{1} \in \real^{round(n/2) \times 1}$, $b_2 \in \real$, $n = |\num| + \sum_{j \in \cat}k_j$ and $\sigma$ is a parametric rectified linear unit (PReLU).

\subsubsection{Graph convolutional network (GCN)}
Considering that an oriented graph is attached to a dataset, meaning that each data point $\xpoint$ is associated to a node $\node{i}$. We denote $\nhood{i}$ the set of nodes $\node{j}$ such that an oriented edge exist from $\node{j}$ to $\node{i}$ and $d_{ji}$ the weight associated to this edge. \\

\noindent Following the Jumping Knowledge framework \cite{Xu:2018}, we implemented the \acrshort{gcn} \cite{Kipf:2017} such that
\begin{equation}
    y_i = \w^{T}\left(\left[\xnum;\xcatprime;\xpoint^{\prime}\right]\right) + b_2
    \label{eq:jknet}
\end{equation}
with
\begin{equation}
    \xpoint^{\prime} = \sigma\left(\frac{1}{\sum_{j \in \nhood{i}}d_{ji}}\sum_{j \in \nhood{i}} d_{ji} \left(\W \left[\mati{m}{j}, \mati{c}{j}^{\prime}\right] + \mati{b}{1}\right) \right)
    \label{eq:gcn_emb}
\end{equation}
where $\W \in \real^{n \times n}$,  $\w \in \real^{2n \times 1}$, $\mati{b}{1} \in \real^{n \times 1}$, $b_2 \in \real$, $n = |\num| + \sum_{j \in \cat}k_j$ and $\sigma$ is a ReLU. \\

\noindent 

\subsubsection{Graph attention network (GAT)}
We designed the \acrshort{gat} \cite{Velickovic:2018} by using the equation (\ref{eq:jknet}) but replacing equation (\ref{eq:gcn_emb}) by 
$$\xpoint^{\prime} = \sigma\left(\sum_{j \in \nhood{i}} \alpha_{ij} \left(\W \left[\mati{m}{j}, \mati{c}{j}^{\prime}\right] + \mati{b}{1}\right) \right)$$
where $\alpha_{ij}$ represents the attention score given to node $\node{j}$ by the node $\node{i}$ and is calculated with a single attention-head of the mechanism introduced by Veli{\v{c}}kovi{\'c} \textit{et al.} \cite{Velickovic:2018}.

\subsubsection{Gene graph attention encoder (GGAE)}
Considering that the set of categorical features is composed by the set of SNPs and the set of categorical features that are not SNPs ($\cat = \snps \cup \clin$). We further define $\snps = \bigcup_{k} \snpsk$ where $\snpsk$ is the set of SNPs belonging in the chromosome pair $k$. \\

\noindent Following this notation, the \acrshort{ggae} first calculates the \textit{chromosomal embeddings} $\mati{z}{i}^{(k)} \in \real^{3\times1}$ linked to a data point $\xpoint$:
$$\mati{z}{i}^{(k)} = \sigma\left(\sum_{j \in \snpsk} \alpha_{ij}^{(k)}\emb \right)$$
with
$$\alpha_{ij}^{(k)} = \frac{exp\left(\gamma\left(\mat{a}^{(k)^T} \emb\right)\right)}{\sum_{p \in \snpsk} exp\left(\gamma\left(\mat{a}^{(k)^T} \matij{h}{i}{p}\right)\right)}$$
where $\mat{a}^{(k)} \in \real^{3\times1}$ is the attention mechanism specific to the chromosome pair $k$, $\emb \in \real^{3\times1}$ is the categorical embedding (i.e., \textit{SNP embedding}) associated to SNPs $j$, $\sigma$ is a ReLU and $\gamma$ is a LeakyReLU. \\

\noindent Then, the \textit{genomic signature} $\mati{s}{i}$ is calculated:
$$\mati{s}{i} = \W\left(\sigma\left(\sum_{k} \beta_{ik} \mati{z}{i}^{(k)} \right)\right)$$
with
$$\beta_{ik} = \frac{exp\left(\gamma\left(\mat{b}^T \mati{z}{i}^{(k)}\right)\right)}{\sum_{\ell} exp\left(\gamma\left(\mat{b}^T \mati{z}{i}^{(\ell)}\right)\right)}$$
where $\W \in \real^{s\times3}$, $\mat{b} \in \real^{3\times1}$ is the final attention mechansim, $s$ is the \textit{signature size}, $\sigma$ is a ReLU and $\gamma$ is a LeakyReLU.

\subsection{Hyperparameters of the models}
\subsubsection{Descriptions of the hyperparameters}
Here, we enumerate and give a description of the hyperparameters associated to the models we implemented. All hyperparameters are associated to the models in which they are included in table \ref{tab:hpsassociations}. For the description of the hyperparameters associated to the random forest and XGBoost, please refer to the documentation of versions 0.24.1 and 1.4.2 of \textit{scikit-learn} \cite{Pedregosa:2011} and \textit{xgboost} \cite{Chen:2016} libraries.

\begin{labeling}{\textbf{number of neighbors}}
\item[attention dropout] Probability of randomly setting an attention score $\alpha_{ij}$ to 0 for any node $\node{i}$ in the execution of a forward pass of the \acrshort{gat} during its training.
\item[batch size] Number of elements in each batch during the training of a model.
\item[$\beta$] L2 penalty coefficient for the regularization term in the mean squared error loss used during the training of a model.
\item[dropout] Probability of randomly setting a feature (i.e., a neuron) of the hidden layer to 0 in the execution of a forward pass of the model during its training.
\item[feature dropout] Probability of randomly setting a feature (i.e., a neuron) of the input layer to 0 in the execution of a forward pass of the \acrshort{gat} during its training.
\item[genomic dropout] Probability of randomly setting a feature within a \textit{SNP embedding} or a \textit{chromosomal embedding} to 0 during the execution of a forward pass of the linear regression with \acrshort{ggae} during its training.
\item[learning rate] Step size parameter $\alpha$ of the Adam optimizer \cite{Kingma:2015}.
\item[number of neighbors] In-degrees of each node in the oriented graph connecting the patients (without considering each node's self-connection).
\item[signature dropout] Probability of randomly setting a feature (i.e., a neuron) of the \textit{genomic signature} to 0 in the execution of a forward pass of the linear regression with \acrshort{ggae} during its training.
\item[signature size] Number of components in the \textit{genomic signature}.
\end{labeling} 

\begin{table}[ht!]
\centering
\begin{tabular}{|c|c|}
\hline
Hyperparameter    & Models                                                                                     \\ \hline \hline
attention dropout & \acrshort{gat}                                                                                        \\ \hline
batch size        & Linear regression, \acrshort{mlp}, Linear regression + \acrshort{ggae} \\ \hline
$\beta$           & All*                                                                                       \\ \hline
dropout           & \acrshort{mlp}, \acrshort{gat}, \acrshort{gcn}, Linear regression + \acrshort{ggae}         \\ \hline
feature dropout   & \acrshort{gat}                                                                                        \\ \hline
genomic dropout   & Linear regression + \acrshort{ggae} \\ \hline
learning rate     & All*                                                                                       \\ \hline
number of neighbors & \acrshort{gat}, \acrshort{gcn} \\ \hline 
signature dropout & Linear regression + \acrshort{ggae}                                                                   \\ \hline
signature size    & Linear regression + \acrshort{ggae}                                                                   \\ \hline
\end{tabular}
\caption[Complete list of the hyperparameters]{List of the hyperparameters and the models to which they are associated. \textbf{*}Models implemented with \textit{Pytorch} and \textit{DGL}.}
\label{tab:hpsassociations}
\end{table}

\subsubsection{Values and search spaces of the hyperparameters}
In the following tables, we report the sets of manually selected hyperparameter values and hyperparameters' search spaces that led to the results obtained for each model. All models implemented with \textit{Pytorch} and \textit{DGL} were trained using the Adam optimizer with a maximal budget of \textbf{500} epochs. The training of each of these models was prematurely stop if no improvement of the root-mean-square error was seen on the validation set for \textbf{50} consecutive epochs. 

\begin{table}[ht!]
\centering
\begin{tabular}{|c|c|c|}
\hline
Hyperparameter   & Manually selected value & Search space                    \\ \hline \hline
max\_features    & sqrt        & $\{ \text{sqrt}, \text{log2}\}$ \\ \hline
max\_leaf\_nodes & 25          & $\{5k\}_{k=1}^{10}$             \\ \hline
max\_samples     & 0.8         & $\left[0.8, 1\right]$           \\ \hline
n\_estimators    & 2000        & $\{1000 + 250k\}_{k=0}^{8}$     \\ \hline
\end{tabular}
\caption[Random forest's hyperparameters.]{Random forest's hyperparameters. The hyperparameters that are not mentioned were set as the default ones from version 0.24.1 of the \textit{scikit-learn} library.}
\label{tab:randomforesthps} 
\end{table}

\begin{table}[ht!]
\centering
\begin{tabular}{|c|c|c|}
\hline
Hyperparameter & Manually selected value                                        & Search space              \\ \hline \hline
max\_depth     & 2                                                  & $\{k\}_{k=1}^{5}$           \\ \hline
learning\_rate & 0.1                                                & $\left[5 \times 10^{-3}, 0.1\right]$ \\ \hline
reg\_lambda    & $5 \times 10^{-4}$ & $\left[5\times 10^{-4}, 1\right]$   \\ \hline
subsample      & 0.8                                               & $\left[0.8, 1\right]$             \\ \hline
\end{tabular}
\caption[XGBoost's hyperparameters.]{XGBoost's hyperparameters. The hyperparameters that are not mentioned were set as the default ones from the scikit-learn wrapper interface of version 1.4.2 of the \textit{xgboost} library.}
\label{tab:xgboosthps}
\end{table}

\begin{table}[ht!]
\centering
\begin{tabular}{|c|c|c|}
\hline
Hyperparameter & Manually selected value        & Search space                          \\ \hline \hline
batch size     & 5                  & $\{5k\}_{k=1}^{5}$                    \\ \hline
$\beta$      & $5 \times 10^{-4}$ & $\left[5 \times 10^{-4}, 1\right]$   \\ \hline
learning rate  & 0.01               & $\left[5 \times 10^{-3}, 0.1\right]$ \\ \hline
\end{tabular}
\caption{Linear regression's hyperparameters.}
\label{tab:linearreghps}
\end{table}

\begin{table}[ht!]
\centering
\begin{tabular}{|c|c|c|}
\hline
Hyperparameter & Manually selected value & Search space                          \\ \hline \hline
batch size     & 5                       & $\{5k\}_{k=1}^{5}$                    \\ \hline
$\beta$           & $5 \times 10^{-4}$      & $\left[5 \times 10^{-4}, 1\right]$    \\ \hline
dropout        & 0.25                    & $[0, 0.25]$                           \\ \hline
learning rate  & $1 \times 10^{-3}$      & $\left[1 \times 10^{-3}, 0.1]\right]$ \\ \hline
\end{tabular}
\caption{\acrshort{mlp}'s hyperparameters.}
\label{tab:mlphps}
\end{table}

\begin{table}[ht!]
\centering
\begin{tabular}{|c|c|c|}
\hline
Hyperparameter & Manually selected value & Search space                          \\ \hline \hline
$\beta$             & $5 \times 10^{-3}$      & $\left[5 \times 10^{-4}, 1\right]$    \\ \hline
dropout        & 0.25                    & 0.25                          \\ \hline
learning rate  & 0.1      & $\left[5 \times 10^{-3}, 0.1\right]$ \\ \hline
number of neighbors & 8 & 8  \\ \hline
\end{tabular}
\caption{\acrshort{gcn}'s hyperparameters.}
\label{tab:gcnhps}
\end{table}

\begin{table}[ht!]
\centering
\begin{tabular}{|c|c|c|}
\hline
Hyperparameter    & Manually selected value & Search space                          \\ \hline \hline
attention dropout & 0.5                     & 0.5                                   \\ \hline
$\beta$           & $5 \times 10^{-3}$      & $\left[5 \times 10^{-4}, 1\right]$    \\ \hline
dropout           & 0.25                    & 0.25                                  \\ \hline
feature dropout   & 0                       & $[0, 0.25]$                           \\ \hline
learning rate     & 0.1                     & $\left[5 \times 10^{-3}, 0.1\right]$ \\ \hline
number of neighbors & $6^{[**]}$, $10^{[*,***]}$ & $6^{[**]}$, $10^{[*,***]}$ \\ \hline
\end{tabular}
\caption[\acrshort{gat}'s hyperparameters]{\acrshort{gat}'s hyperparameters. *: Value selected for the \vo peak problem. **: Value selected for the early obesity problem (w/o SNPs). ***: Value selected for the early obesity problem (w/ SNPs)}
\label{tab:gathps}
\end{table}

\begin{table}[ht!]
\centering
\begin{tabular}{|c|c|c|}
\hline
Hyperparameter    & Manually selected value & Search space                          \\ \hline \hline
batch size        & 5                       & $\{5k\}_{k=1}^{5}$                    \\ \hline
$\beta$           & $5 \times 10^{-4}$      & $\left[5 \times 10^{-4}, 1\right]$    \\ \hline
genomic dropout   & 0.25                    & [0, 0.25] \\ \hline
learning rate     & 0.1                     & $\left[5 \times 10^{-3}, 0.1]\right]$ \\ \hline
signature dropout & 0.25                    & 0.25                                  \\ \hline
signature size    & 4                       & 4                                     \\ \hline
\end{tabular}
\caption{Linear regression + \acrshort{ggae}'s hyperparameters.}
\label{tab:ggaehps}
\end{table}

\clearpage

\subsection{Tree-structured Parzen estimator (TPE)}
\begin{figure}[ht!]
    \centering
    \includegraphics[scale=0.60]{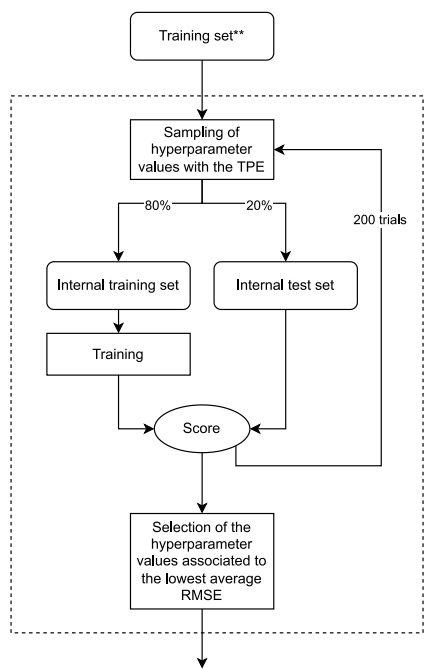}
    \caption[Automated hyperparameter optimization with the Tree-Structured Parzen estimator algorithm (TPE).]{Automated hyperparameter optimization with the Tree-Structured Parzen estimator algorithm (TPE). 200 sets of hyperparameter values are sampled sequentially using the TPE and evaluated on the 10 same \textit{internal train sets} and \textit{internal test sets}. The set of hyperparameter values associated to the lowest average RMSE is selected. The dashed rectangle refers to the feature selection steps of boxes 2 and 4 in figure 1\textbf{b} of the main text. \textbf{**}: training set after the feature selection. }
    \label{fig:automated_hps_optim}
\end{figure}

In this subsection, we expose additional details about the parameters (Supplementary Table \ref{tab:tpe_param}) and the statistical distributions (Supplementary Table \ref{tab:distributions}) used to run the hyperparameter optimization with the TPE algorithm \cite{Bergstra:2011}. Additionally, an illustration of the automated hyperparameter optimization process is displayed (Supplementary Fig. \ref{fig:automated_hps_optim})

\begin{table}[ht!]
\centering
\begin{tabular}{|c|c|}
\hline
Parameter                       & Value          \\ \hline \hline
Expected improvement candidates & 20             \\ \hline
Startup trials                  & 20             \\ \hline
\end{tabular}
\caption{Parameters configuration for the TPE algorithm.}
\label{tab:tpe_param}
\end{table}

\begin{table}[ht!]
\centering
\begin{tabular}{|c|c|}
\hline
Type &        Distribution            \\ \hline \hline
Integer ($\mathbb{Z}$)       & Integer Uniform       \\ \hline
Real ($\mathbb{R}$)    & Uniform               \\ \hline
\end{tabular}
\caption[Statistical distributions for the TPE algorithm]{Statistical distributions used to sample hyperparameters according to their types.}
\label{tab:distributions}
\end{table}

\section{SUPPLEMENTARY BACKGROUND}
\subsection{Disease-specific \vo peak equation}
Here, we present the disease-specific \vo peak equation created by Labonté \textit{et al.} \cite{Labonte:2020}.
\begin{align*}
    VO_2 \ peak \ (mL/kg/min) \ &= -0.236\times Age \ (years) - 0.094\times Weight \ (kg) - 0.120\times \acrshort{hrend} \ (bpm) \\
                & \ + 0.067\times \acrshort{6mwd} \ (m) + 0.065\times \acrshort{mvlpa} \ (min/day) - 0.204\times \acrshort{dt} \ (years) + 25.145
\end{align*}
\newpage
\printglossary[title=Abbreviations, toctitle=Abbreviations]
\newpage

\printbibliography